\documentclass[12pt,a4paper]{article}
\pdfoutput=1
\usepackage{jheppub}
\usepackage[sort&compress]{natbib}
\usepackage{url}
\usepackage{hyperref}
\usepackage{amsfonts}
\usepackage{epsfig}\usepackage{dcolumn}
\usepackage{bm}
\usepackage{slashed}
\graphicspath{{./figuras/}}
\usepackage{subfigure}

\newcommand{\zp}{Z^{\prime}}

\title{{\boldmath{\color{blue} Explaining Dark Matter and $B$ Decay Anomalies with an $L_\mu - L_\tau$ Model}}}

\author{Wolfgang~Altmannshofer$^{a}$,}
\author{Stefania~Gori$^{a}$,}
\author{Stefano~Profumo$^{b}$,}
\author{Farinaldo~S.~Queiroz$^{c}$}

\affiliation{$^a$Department of Physics, University of Cincinnati, Cincinnati, Ohio 45221, USA}
\affiliation{$^b$Department of Physics and Santa Cruz Institute for Particle Physics
University of California, Santa Cruz, CA 95064, USA}
\affiliation{$^c$ Max-Planck-Institut fur Kernphysik,
Saupfercheckweg 1, 69117 Heidelberg, Germany}

\emailAdd{altmanwg@ucmail.uc.edu}
\emailAdd{stefania.gori@uc.edu}
\emailAdd{profumo@ucsc.edu}
\emailAdd{farinaldo.queiroz@mpi-hd.mpg.de}

\abstract{We present a dark sector model based on gauging the $L_\mu - L_\tau$ symmetry that addresses anomalies in $b \rightarrow s \mu^+ \mu^-$ decays and that features a particle dark matter candidate.
The dark matter particle candidate is a vector-like Dirac fermion coupled to the $Z^\prime$ gauge boson of the $L_{\mu}-L_{\tau}$ symmetry. We compute the dark matter thermal relic density, its pair-annihilation cross section, and the loop-suppressed dark matter-nucleon scattering cross section, and compare our predictions with current and future experimental results.
We demonstrate that after taking into account bounds from $B_s$ meson oscillations, dark matter direct detection, and the CMB, the model is highly predictive: $B$ physics anomalies and a viable particle dark matter candidate, with a mass of $\sim (5-23)$~GeV, can be accommodated only in a tightly-constrained region of parameter space, with sharp predictions for future experimental tests. The viable region of parameter space expands if the dark matter is allowed to have $L_\mu-L_\tau$ charges that are smaller than those of the SM leptons.}

\begin{document} 
\maketitle
\flushbottom

\section{Introduction}

The presence of non-baryonic dark matter in the universe has been ascertained from a variety of observations corresponding to different time scales in the history of the universe. All such observations hinge, however, exclusively on gravitational effects associated with the dark matter, and no striking evidence for non-gravitational manifestations of dark matter is known to date.  A broad experimental program is presently in place, aimed at discovering such non-gravitational effects with collider searches and with direct and indirect dark matter detection.

Collider searches for dark matter typically rely on so-called mono-$X$ searches, i.e. missing energy plus some visible final state (see~\cite{Aaboud:2016qgg,ATLAS:2016tsc,Aaboud:2016tnv,Aaboud:2016uro,monoHiggsATLAS,CMS:2016flr,CMS:2016fnh,CMS:2016hmx,CMS:2016pod,CMS:2016mjh,CMS:2016xok,CMS:2016mxc,CMS:2016uxr} for the latest analyses); such searches have the advantage of being sensitive to low dark matter masses, as opposed to e.g. current direct detection experiments which loose sensitivity for small masses below a few GeV. In addition to the usual mono-$X$ studies, colliders might also give important information on the mediator that connects the dark and visible sector, for example in the case of vector mediators, as we will describe in more detail below.

In the class of models we investigate here, the dark matter is connected to Standard Model (SM) particles via a new gauge interaction that is also relevant to address putative new physics (NP) signals observed in rare $B$ meson decays by the LHCb collaboration. 

Specifically, we build on the model presented  in Refs.~\cite{Altmannshofer:2014cfa,Altmannshofer:2015mqa}, which is based on gauging $L_{\mu}-L_{\tau}$, the difference of leptonic muon number and tau number~\cite{He:1990pn,He:1991qd,Heeck:2011wj}. The model of~\cite{Altmannshofer:2014cfa,Altmannshofer:2015mqa} was proposed to explain an anomaly in the rare $B \rightarrow K^{\star} \mu^+ \mu^-$ decay mode observed by the LHCb collaboration in 2013~\cite{Aaij:2013qta}.\footnote{See \cite{Glashow:2014iga,Crivellin:2015mga,Varzielas:2015iva,Crivellin:2015lwa,Niehoff:2015bfa,Sierra:2015fma,Celis:2015ara, Greljo:2015mma,Belanger:2015nma,Falkowski:2015zwa,Allanach:2015gkd,Fuyuto:2015gmk,Becirevic:2016zri,Altmannshofer:2016oaq,Boucenna:2016qad,Megias:2016bde} for related NP explanations of $B$ physics anomalies.}
The $B \rightarrow K^{\star} \mu^+ \mu^-$ anomaly persists in the latest experimental data~\cite{Aaij:2015oid,Abdesselam:2016llu}, and is supported by measurements of the $B_s \rightarrow \phi \mu^+ \mu^-$~\cite{Aaij:2015esa} decay and by the hint for lepton flavor universality violation in $B^+ \to K^+ \ell^+ \ell^-$~\cite{Aaij:2014ora}.
Interestingly enough, the model in~\cite{Altmannshofer:2014cfa,Altmannshofer:2015mqa} necessarily predicts lepton flavor universality violation and is in excellent agreement with the latest model independent NP fits that take into account all relevant experimental data on rare $B$ decays~\cite{Altmannshofer:2014rta,Altmannshofer:2015sma,Descotes-Genon:2015uva,Hurth:2016fbr}.

Beyond being able to explain the hints for lepton flavor universality violation in rare $B$ decays, gauged $L_\mu - L_\tau$ has a number of additional virtues. It has been considered as a solution to the anomaly in the $g-2$ of the muon, and is used in models for neutrino masses. Unbroken $L_\mu - L_\tau$ predicts one degenerate neutrino pair and a maximal atmospheric neutrino mixing angle, which is a good starting point for model building.

Extra matter could be charged under the $L_\mu - L_\tau$ gauge symmetry. Here, we consider a minimal dark sector composed by the $Z^\prime$ gauge boson associated to the $L_\mu-L_\tau$ symmetry and by a vector-like Dirac fermion that is a dark matter candidate, and which we assume to be charged under the new local $L_{\mu}-L_{\tau}$ symmetry. We investigate whether both the rare $B$ decay anomalies and dark matter can be simultaneously accounted for, considering all relevant phenomenological constraints. The dark matter phenomenology is dictated by gauge interactions mediated by the massive new $Z^{\prime}$ gauge boson. The mass of the $Z^{\prime}$ can be generated either through a Stueckelberg mechanism, thereby leaving the $L_{\mu}-L_{\tau}$ symmetry unbroken~\cite{Ruegg:2003ps,Feldman:2006wb}, or by a spontaneous symmetry breaking mechanism with the introduction of a scalar field, as described in Ref.~\cite{Altmannshofer:2014cfa}. We will assume that the scalar field, if it exists, is either sufficiently heavy to be integrated out and inessential to the phenomenology of the model, or that is has sufficiently suppressed interactions with SM particles, again leading to no observable consequences. In what follows we remain agnostic about the mechanism by which the $Z^{\prime}$ gauge boson acquires mass, and we outline the region of parameter space which accommodates both dark matter and the rare $B$ decay anomalies.

Notice that the construction under consideration here is primarily motivated by the notion of minimality in addressing both dark matter and B physics anomalies. As a result, while a full underlying model might reveal a connection between the ingredients of the model (e.g. the vector-like dark matter particle and the vector-like quarks that generate the effective couplings between the $Z\prime$ and the Standard Model quarks), none is postulated here. The construction of a complete model, including connections among neutrino masses, dark matter and vector-like quarks goes beyond the scope of this work.

Previous work has explored dark matter in models based on the $L_{\mu}-L_{\tau}$ symmetry. 
Early studies of dark matter charged under $L_\mu - L_\tau$ were motivated by the PAMELA positron fraction excess and the anomalous magnetic moment of the muon~\cite{Baek:2008nz,Bi:2009uj,Queiroz:2014zfa} and the galactic center excess~\cite{Kim:2015fpa}. Extended $L_{\mu}-L_{\tau}$ setups have also been explored, containing right-handed neutrinos and additional light scalars~\cite{Baek:2015fea} or additional gauge bosons~\cite{Kile:2014jea}.
More studies of $L_{\mu}-L_{\tau}$ models with extended fermion and scalar sectors have been performed in~\cite{Patra:2016shz,Biswas:2016yan}. A complementary study of the interplay of $B$ physics anomalies and Dark Matter in models with gauged Baryon number can be found in~\cite{Celis:2016ayl}.

Our work extends and complements
previous studies in several notable directions:
\begin{itemize}
\item We postulate a Dirac fermion as a dark matter candidate and discuss several bounds on the $Z^{\prime}$ gauge boson stemming from neutrino trident production and collider searches in the context of a local $L_{\mu}-L_{\tau}$ symmetry.
\item We consider stringent limits stemming from indirect detection, specifically from CMB measurements on the energy injection at dark ages. 
\item We account for 1-loop dark matter-nucleon spin-independent scattering, and compare such predictions with the LUX and projected XENON1T experimental sensitivity. Despite the 1-loop suppression, we find remarkably strong constraints from direct detection experiments.
\item Lastly, we outline the region of parameter space where both dark matter and the rare $B$ decay anomalies can be simultaneously addressed in agreement with all existing bounds. 
\end{itemize}

The paper is organized as follows: Sec.~\ref{sec:model} introduces the model under consideration; Sec.~\ref{sec:constraints} discusses general constraints on the $Z^\prime$ parameter space, while the following Sec.~\ref{sec:anomalies} details on the connection with the $B$ decay anomalies and Sec.~\ref{sec:dm} discusses the dark matter-related phenomenology. Finally, Sec.~\ref{sec:global} describes and summarizes our findings, and Sec.~\ref{sec:conclusions} concludes.

\section{\boldmath The $L_\mu - L_\tau$ Model}\label{sec:model}

In the $L_\mu - L_\tau$ model, the SM is augmented by a new Abelian symmetry, $U(1)_{\mu-\tau}$, with a corresponding new massive gauge boson, $Z^{\prime}$. We remain agnostic as to how the $Z^\prime$ acquires its mass,
but we do assume that the physics connected to the mass generation is sufficiently decoupled and phenomenologically irrelevant for our discussion. The new boson couples to the SM fields via the covariant derivative $D_{\alpha} =\partial_{\alpha} + ig^{\prime} q_{\mu-\tau} Z^{\prime}$~, with $g^{\prime}$ the $U(1)_{\mu-\tau}$ coupling strength and $q_{\mu-\tau}$ the corresponding charge, according to the neutral current
\begin{equation}
{\cal L}_{\rm fermions} \supset q_\ell g^{\prime} \left( \bar{\ell}_2 \gamma_{\alpha} \ell_2 - \bar{\ell}_3 \gamma_{\alpha} \ell_3 + \bar{\mu}_R \gamma_{\alpha} \mu_R - \bar{\tau}_R \gamma_{\alpha} \tau_R \right) Z^{\prime\alpha} \,,
\end{equation}
where $q_\ell$ is a free parameter which quantifies the overall charge of the leptons under the $L_\mu - L_\tau$ symmetry. 
$\ell_i$ indicates the SM lepton doublets with flavor $i$. This interaction results into,
\begin{equation}
{\cal L}_{\rm fermions} \supset q_\ell g^{\prime} \left( \bar{\mu} \gamma_{\alpha} \mu - \bar{\tau} \gamma_{\alpha} \tau + \bar{\nu_{\mu}} \gamma_{\alpha} P_L \nu_{\mu} - \bar{\nu_{\tau}} \gamma_{\alpha} P_L \nu_{\tau} \right)Z^{\prime\alpha} \,.
\label{model:eq1}
\end{equation}
Dark matter can be charged under this new gauge symmetry. A simple extension of the minimal model that incorporates a dark matter candidate is constructed by adding a vector-like Dirac fermion $\chi$, singlet under the SM gauge group but charged under the new $U(1)_{\mu-\tau}$ symmetry. The fermion has a vector-like mass, $m_\chi$, and its coupling to the $Z^\prime$ is given by
\begin{equation}
{\cal L}_\text{dark} \supset  q_{\chi}\,  g^{\prime}\, \bar{\chi}\gamma_{\alpha} \chi Z^{\prime \alpha} \,,
\label{model:eq2}
\end{equation}
where $q_{\chi}$ is the dark matter charge under the $U(1)_{\mu-\tau}$ symmetry. 
The vector-like nature of the dark matter particle ensures the absence of triangle anomalies. In our analytic results we keep the dependence on the two charges $q_\ell$ and $q_{\chi}$ explicit. For our numerical results, however, we adopt $q_\ell=1$, without loss of generality. The larger the $q_\ell$ $L_\mu - L_\tau$ charge, the smaller the $g^{\prime}$ value needed to reproduce the same results. We will mainly concentrate on the case $q_\chi = q_\ell$, i.e. we assume that the Dirac fermion dark matter candidate features a ``universal coupling'' to the $Z^{\prime}$, equal to that of leptons. We will also comment in what follows on the potentially phenomenologically interesting case $q_\chi \ll q_\ell$.

\section{\boldmath Constraints on the $Z^\prime$ Parameter Space}\label{sec:constraints}

A powerful probe of a $Z^\prime$ vector boson based on gauging $L_\mu - L_\tau$ is the process of neutrino trident production~\cite{Altmannshofer:2014pba}, i.e. the production of a $\mu^+\mu^-$ pair in the scattering of a muon neutrino in the Coulomb field of a heavy nucleus.
Integrating out the $Z^\prime$, the correction to the trident cross section can be concisely written as~\cite{Altmannshofer:2014cfa} 
\begin{equation} \label{eq:trident}
 \frac{\sigma_{\text{SM}+Z^\prime}}{\sigma_\text{SM}} = \frac{1 + (1+4 s_W^2 + 2 v^2 q_\ell^2 (g^\prime)^2/m_{Z^\prime}^2)^2}{1+(1+4s_W^2)^2}\,,
\end{equation}
where $v = 246$~GeV is the electroweak vacuum expectation value (vev) and $s_W \equiv \sin\theta_W$ is the sine of the Weinberg angle $\theta_W$. Using the measurement of the trident cross section by the CCFR collaboration~\cite{Mishra:1991bv}, 
\begin{equation}
\sigma_\text{CCFR}/\sigma_\text{SM} = 0.82 \pm 0.28 \,,
\end{equation}
one finds, for a given $g^\prime$ gauge coupling, the following lower bound on the $Z^\prime$ mass 
\begin{equation} \label{eq:tridentbound}
m_{Z^\prime} > 540~\text{GeV} \times q_\ell\times  g^\prime \,.
\end{equation}
This bound is compatible with very light $Z^\prime$ bosons as long as the coupling $g^\prime$ is very small. For a light $Z^\prime$, with mass comparable to the neutrino momentum transfer in the trident reaction, the approximate expression in~(\ref{eq:trident}) breaks down. In this region of parameter space the $Z^\prime$ has to be kept as dynamical degree of freedom. For $m_{Z^\prime} \lesssim 10$~GeV we use the results of a numerical evaluation of the trident cross section in the presence of a light $Z^\prime$ from Ref.~\cite{Altmannshofer:2014pba}. For such light $Z^\prime$ the constraint in~(\ref{eq:tridentbound}) is slightly relaxed.
The region that is excluded by the trident measurements is shaded in red in Figs.~\ref{queirozplot01} - \ref{queirozplot6}. We stress that this constraint is independent of the DM parameter space as defined by the parameters ($q_\chi, m_\chi$).

An additional constraint on the parameter space of the $L_\mu - L_\tau$ gauge boson comes from measurements of the $Z\to 4\ell$ branching ratio~\cite{Altmannshofer:2014cfa}. 
The ATLAS collaboration has measured \cite{TheATLAScollaboration:2013nha} the fiducial branching ratio
\begin{equation} \label{eq:Z4l}
{\rm BR}(Z\to 4\ell)=(4.2\pm 0.4)\times 10^{-6} \,,
\end{equation}
using the full 7 and 8~TeV data set. This measurement is in good agreement with the SM prediction 
\begin{equation}
{\rm BR}(Z\to 4\ell)_{\rm SM}=(4.37\pm 0.03)\times 10^{-6} \,.
\end{equation}
An additional ATLAS Run I analysis~\cite{Aad:2014wra} leads to very similar results. Also the CMS collaboration has performed a search for $Z\to 4\ell$ using 2.6~fb$^{-1}$ of 13~TeV data~\cite{CMS:2016vvl}. The 13~TeV result, however, has a still larger (statistical) uncertainty, if compared to the ATLAS search based on the full 8 TeV data set.
We do not attempt a statistical combination of these results, but will use the measurement~(\ref{eq:Z4l}) in the following.
Our $Z^\prime$ contributes to the process with the 4 muon final state. 
Despite the fact that the ATLAS search has not been optimized to specifically constrain the $Z^\prime$ scenario, interesting bounds on the gauge coupling $g^\prime$ can be set for $4\gtrsim m_{Z^\prime}/{\rm GeV}\gtrsim 70$ where the three-body decay $Z \to \mu^+\mu^- Z^\prime$ is open. For larger $Z^\prime$ masses the phase space starts to close, while for lower $Z^\prime$ masses, the experimental acceptance becomes too small: the ATLAS analysis requires, in fact, two independent pairs of leptons with invariant masses above 5 GeV, implying that in order to get a non zero acceptance for $m_{Z^\prime}< 5$~GeV, one would need to ``mispair'' the leptons, since, otherwise, one pair would have an invariant mass $m_{\mu\mu}=m_{Z^\prime}<5$ GeV. Secondly, going to lower and lower $Z^\prime$ masses, the two leptons coming from the $Z^\prime$ decay become more and more collimated, failing more often the isolation criterium employed in~\cite{TheATLAScollaboration:2013nha}.

The region in the $m_{Z^\prime}$ - $g^\prime$ plane that has been probed by the $Z \to 4 \ell$ measurement is shown in gray, in Figs.~\ref{queirozplot01} - \ref{queirozplot6}. The most stringent limit on the gauge coupling, $g^\prime \lesssim 0.015$, can be set at around the mass $m_{Z^\prime} \simeq 10$ GeV. It might be possible to substantially extend the collider reach in the coming years of the LHC using targeted new searches for $Z\to 2\mu Z^\prime$, $Z^\prime \to 2\mu$ and for $Z\to 2\mu Z^\prime,~Z^\prime\to\nu\nu$~\cite{Elahi:2015vzh}. Notice that the LHC bound on the $Z^\prime$ depends slightly on the DM mass: if the $Z^\prime$ is heavier than two times the DM, then the $Z^\prime$ will also decay to a DM pair, weakening in such a way the bound coming from the measurement of $Z \to 4 \ell$. We take into account this effect in our numerical results.

Finally, we point out that the BaBar collaboration has recently performed a search for muonic forces measuring the cross section for the process
\begin{equation}
e^+ e^- \to \mu^+\mu^- Z^\prime \,,
\end{equation}
with the $Z^\prime$ decaying into $\mu^+\mu^-$~\cite{TheBABAR:2016rlg}. The search constrains regions of $Z^\prime$ parameter space with $2 m_\mu < m_{Z^\prime} \lesssim 8$~GeV. The corresponding constraints on $g^\prime$ are slightly stronger than the trident constraints for $m_{Z^\prime} \lesssim 4$~GeV. Such low $Z^\prime$ masses are outside of our main region of interest. Therefore the resulting bound will not be presented in what follows.

\section{Rare B Decay Anomalies}\label{sec:anomalies}

Over the last few years, various anomalies in rare $B$ meson decays have been reported by the LHCb Collaboration \cite{Blake:2016olu}. Such anomalies include discrepancies in the $B \to K^* \mu^+\mu^-$ angular distribution~\cite{Aaij:2013qta,Aaij:2015oid} (confirmed by a recent Belle result~\cite{Abdesselam:2016llu}), a reduced branching ratio for the decay mode $B_s \to \phi \mu^+\mu^-$~\cite{Aaij:2015esa}, as well as a hint for lepton flavor universality violation in $B \to K \mu^+\mu^-$ vs. $B \to K e^+e^-$~\cite{Aaij:2014ora}. Assuming that hadronic uncertainties in the corresponding SM predictions are estimated in a sufficiently conservative way, global fits of the combined rare $B$ meson decay data show a strong preference for new physics in $b \to s \mu\mu$ transitions, while $b \to s ee$ transitions appear compatible to the corresponding SM predictions~\cite{Altmannshofer:2014rta,Altmannshofer:2015sma,Descotes-Genon:2015uva,Hurth:2016fbr}.
The best description of the data is obtained by new physics in the form of a four fermion contact interaction 
\begin{equation} \label{eq:Heff}
\mathcal{H}_\text{eff}^\text{NP} = -\frac{4 G_F}{ \sqrt{2}} \frac{\alpha_{em}}{4\pi} (V_{tb} V_{ts}^*) ~C_9^\text{NP} (\bar s \gamma_\alpha P_L b)(\bar \mu \gamma^\alpha \mu) ~,
\end{equation}
with $C_9^\text{NP}= -1.07$ and $V$ is the CKM matrix. 
The $Z^\prime$ arising from the $L_\mu - L_\tau$ gauge symmetry is ideally suited to address the $B$ physics anomalies as it has the required vector couplings to muons and does not couple, by construction, to electrons. As shown in Ref.~\cite{Altmannshofer:2014cfa}, introducing effective flavor-changing couplings of the $Z^\prime$ gauge boson to left handed quarks and integrating out the $Z^\prime$ leads precisely to the contact interaction in Eq.~(\ref{eq:Heff}). 

The Wilson coefficient $C_9^\text{NP}$ is determined by three parameters: the $Z^\prime$ mass, $m_{Z^\prime}$, the $Z^\prime$ coupling to muons, $ q_\ell g^\prime$, and its flavor violating $b\leftrightarrow s$ coupling. In the following we choose $m_{Z^\prime}$ and $g^\prime$ as free parameters and set the flavor-violating coupling such that the rare $B$ decay anomalies are explained, i.e. such that the best fit value for $C_9^\text{NP} = -1.07$ from~\cite{Altmannshofer:2015sma} is reproduced. 
This is possible as long as the $Z^\prime$ boson is sufficiently heavy compared to the $B$ mesons. 
A $m_{Z^\prime}$ below the mass of $B$ mesons would lead, in fact, to a resonance in the dimuon invariant mass spectrum in rare $B$ decays 
which is strongly constrained~\cite{Aaij:2015tna}.
To ensure a consistent explanation of the $B$ anomalies we will demand the conservative bound $m_{Z^\prime} \gtrsim 10$~GeV. 

Note that for such $Z^\prime$ masses, the effective operator approach of eq.~(\ref{eq:Heff}) is fully justified. The region of di-muon invariant mass squared that is most relevant for the rare $B$ decay anomalies is $1~\text{GeV}^2  <  q^2  < 6~\text{GeV}^2$. For $m_{Z^\prime} \gtrsim 10$~GeV, corrections are therefore at the few percent level at most.

Any explanation of the $B$ decay anomalies based on $Z^\prime$ bosons is subject to additional constraints from $B_s$ meson oscillations. Indeed, integrating out the $Z^\prime$ does not only lead to contributions to the $b\to s \mu^+\mu^-$ decays. Once the flavor violating $b \leftrightarrow s$ coupling is fixed to explain the anomalies, it necessarily leads also to corrections to $B_s$ meson oscillations. We find the following modification of the $B_s - \bar B_s$ mixing amplitude $M_{12}$
\begin{equation}
\frac{M_{12}^{Z^\prime}}{M_{12}^\text{SM}} \simeq \frac{m_{Z^\prime}^2}{q_\ell^2 (g^\prime)^2 v^2} s_W^2 \frac{\alpha_{em}}{\pi} \frac{|C_9^\text{NP}|^2}{S_0} \,,
\end{equation}
with the SM loop function $S_0 \simeq 2.3$~\cite{Buchalla:1995vs}. 
Allowing for $|M_{12}^{Z^\prime}|/|M_{12}^\text{SM}| \lesssim 15\%$~\cite{Charles:2015gya,Bazavov:2016nty} and setting $C_9^\text{NP} = -1.07$~\cite{Altmannshofer:2015sma}, as favored by the anomalies in $B$ meson decays, we obtain the following {\it upper} bound on the $Z^\prime$ mass
\begin{equation}
 m_{Z^\prime} < 4.9~\text{TeV} \times q_\ell g^\prime \times \left(\frac{-1.07}{C_9^\text{NP}}\right) \times \left(\frac{|M_{12}^{Z^\prime}|/|M_{12}^\text{SM}|}{15\%}\right)^{\frac{1}{2}} \,.
\end{equation}
This bound in shown in Figs.~\ref{queirozplot01} - \ref{queirozplot6} in magenta.

In summary, the parameter space that allows to address the $B$ decay anomalies, while being consistent with the trident and $Z\to 4\ell$ constraints, is characterized by a lower bound on the ratio $m_{Z^\prime}/( q_\ell g^\prime)$ coming from neutrino tridents, an upper bound on $m_{Z^\prime}/( q_\ell g^\prime) $ coming from $B_s$ meson oscillations and a lower bound on $m_{Z^\prime}$ from the dimuon invariant mass distribution in rare $B$ decays
\begin{equation}
 540~\text{GeV} \lesssim m_{Z^\prime}/ (q_\ell g^\prime) \lesssim 4.9~\text{TeV} ~,~~ m_{Z^\prime} \gtrsim 10~\text{GeV} ~. 
\end{equation}
Outside this window of favored parameter space, the model cannot address the $B$ decay anomalies. We conclude this section noting that this parameter region is completely independent of the DM parameter space $(q_\chi, m_\chi)$.

\section{Dark Matter Phenomenology}\label{sec:dm}

\begin{figure}[!t]
\centering
\includegraphics[scale=0.4]{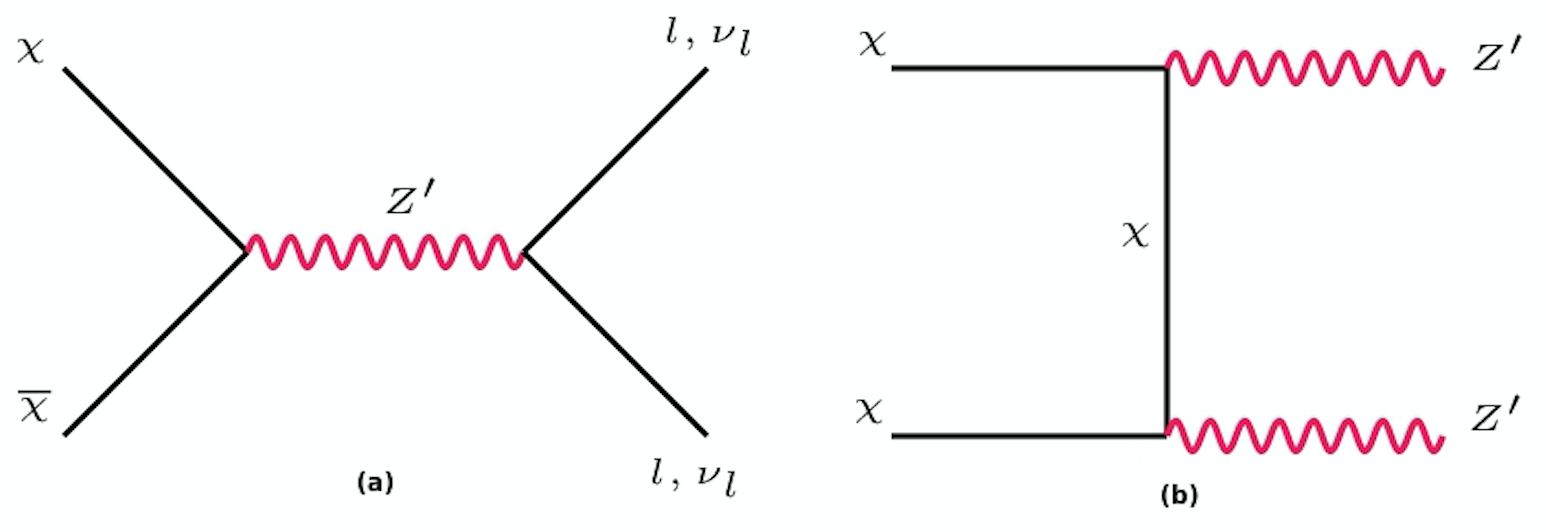}
\caption{Main Dark Matter annihilation channels. For process (a) we have $l = \mu, \tau$. Process (b) is relevant only if kinematically allowed, i.e. when $m_{\chi} > m_{Z^{\prime}}$.}
\label{fig01}
\end{figure}

In this section we discuss the phenomenology of the dark matter in our  model.  It is important to first notice  that the $Z^{\prime}$ gauge boson arising from the $L_{\mu}-L_{\tau}$ local symmetry dictates a dark matter phenomenology rather different from other realizations of a ``dark $Z^{\prime}$ portal''
(see e.g. \cite{Mizukoshi:2010ky,Profumo:2013sca, Alves:2013tqa, Arcadi:2014lta, Buchmueller:2014yoa, Cline:2014dwa, Fairbairn:2014aqa, Lebedev:2014bba, deSimone:2014pda, Okada:2016gsh, Alves:2015mua, Kahlhoefer:2015bea, Duerr:2015wfa, Brennan:2016xjh, Jacques:2016dqz, Englert:2016joy, D'Eramo:2016atc,Klasen:2016qux}). For example, the dark matter pair annihilation exclusively yields muon, tau, and neutrino pairs, as well as $Z^{\prime}$ boson pairs, if kinematically allowed (i.e. if $m_\chi>m_{Z^\prime}$). We show the schematic relevant Feynman diagrams  in Fig. \ref{fig01}. This feature affects both the relic density and the indirect detection properties of the model. Secondly, scattering off of nuclei (direct detection) occurs only via loop induced dark matter-nucleon interactions (see Fig.~\ref{figure:DD}). Most of the previous results obtained in the literature are therefore not directly applicable here.

\subsection{Relic Density}

In the early universe, the dark matter particle was in thermal equilibrium with  SM particles through the interactions shown in Fig. \ref{fig01}. As the universe expanded and cooled down, the expansion rate eventually became equal to the interaction rate, leading to the freeze-out of the dark matter particles, and to a relic dark matter population. Since in our model the interactions of the dark matter particles with  SM particles are dictated by $s$-channel (velocity-independent) processes, away from the $m_\chi\simeq m_{Z^\prime}/2$ resonance, the abundance is directly tied to the annihilation times relative velocity of the annihilating DM particles today. In the case of charged leptons in the final state, such pair-annihilation cross section times relative velocity is given by
\begin{equation}\label{eqannihilation}
\sigma v \left( \chi \bar{\chi} \to \ell^+ \ell^- \right) \approx \frac{ q_{\chi}^2 q_{\ell}^2 g^{\prime 4} }{2 \pi} \sqrt{1-\frac{m_\ell^2}{m_\chi^2}} \frac{2 m_\chi^2 + m_\ell^2}{\left( 4 m_\chi^2 - m_{\zp}^2 \right)^2}\,,
\end{equation}
where $\ell=\mu,\tau$. In the case of neutrinos in the final state the expression is very similar
\begin{equation}\label{eqannihilation2}
\sigma v \left( \chi \bar{\chi} \to \nu \bar\nu \right) \approx \frac{ q_{\chi}^2 q_{\ell}^2 g^{\prime 4} }{2 \pi} \frac{m_\chi^2}{\left( 4 m_\chi^2 - m_{\zp}^2 \right)^2}\,,
\end{equation}
for each neutrino flavor $\nu = \nu_{\mu},\nu_{\tau}$. 
Finally, in the case of final state gauge bosons $Z^{\prime}$,
\begin{align}
\label{equation:cascade}
\sigma v \left( \chi \bar{\chi} \to \zp \zp \right) &\approx \frac{g^{\prime 4} q_{\chi}^4}{16 \pi m_\chi^2} \left(1 - \frac{m_{\zp}^2}{m_\chi^2}\right)^{3/2} \left(1 - \frac{m_{\zp}^2}{2 m_\chi^2}\right)^{-2} \,, 
\end{align} 
relevant when $m_{\chi} > m_{Z^{\prime}}$. 
The quasi-on-shell $Z^\prime$ exchange for $m_\chi\simeq m_{Z^\prime}/2$ gives rise to a resonant enhancement of the cross section. 
In this region of parameter space, Eqs.~(\ref{eqannihilation}) and~(\ref{eqannihilation2}) have to be changed taking into account the $Z^\prime$ width. In our numerical analysis this is taken into account using the {\tt micrOMEGAs} code~\cite{Belanger:2006is,Belanger:2008sj}. From Eqs.~(\ref{eqannihilation})~-~(\ref{equation:cascade}) we learn that 
annihilation into neutrinos and charged leptons occurs at similar rates, and is also comparable to the annihilation rate into $Z^{\prime}$ gauge boson pairs if $m_{\chi} > m_{Z^{\prime}}$. 

Given the annihilation cross section above, it is straightforward to obtain the dark matter thermal relic density, which is approximately
\begin{equation}
\Omega_{\chi} h^2 \simeq \frac{1.04 \times 10^9 x_F{\rm{GeV}}^{-1}}{\sqrt{g_{\star}}M_{\rm{pl}} (\sigma v)} \,,
\end{equation}
with the Planck mass $M_{\rm pl}=1.22\times 10^{19}$ GeV, $x_F $ the inverse temperature at freeze-out in units of the dark matter particle mass, and $g_{\star}$  the number of relativistic degrees of freedom at freeze-out (around 90 for freeze out temperatures of (5-80) GeV \cite{Beltran:2008xg,Beltran:2010ww}). Notice that for the actual numerical evaluation of the dark matter thermal relic abundance we use the {\tt micrOMEGAs} code~\cite{Belanger:2006is,Belanger:2008sj}. 

In Figs.~\ref{queirozplot01} - \ref{queirozplot6} we show in green the curves that reproduce the observed thermal relic abundance $\Omega_\chi h^2 \simeq 0.12$~\cite{Ade:2015xua} as a function of the $g^\prime$ coupling and the $Z^\prime$ mass, for a given value of the particle dark matter mass.  Above these curves our DM candidate is under-abundant. The figures illustrate that there are three distinct regimes: 
\begin{itemize}
\item[(i)] for $m_\chi\gg m_{Z^\prime}$, the relevant annihilation cross section scales as the combination $g^{\prime4}/m_\chi^2$ (for given choices of $q_{\chi},\ q_f$): the correct relic density thus singles out one value of $g^\prime$ for a given dark matter particle mass $m_\chi$;
\item[(ii)] for $m_\chi\simeq m_{Z^\prime}/2$ the resonant regime sets in, driving the $g^\prime$ producing the right relic density to suppressed values;
\item[(iii)] for  $m_\chi\ll m_{Z^\prime}$ the cross section scales as the combination $g^{\prime4}m_\chi^2/m_{Z^\prime}^4$, and thus the right relic density selects a value for the ratio $g^\prime/m_{Z^\prime}$, for a given value of $m_\chi$.
\end{itemize}

The regions of correct thermal relic density on the plane defined by the $g^\prime$ coupling versus the dark matter mass for a given $Z^\prime$ mass are shown in green in Fig.~\ref{queirozplot7}. The main features of the curves can be again simply understood from the approximate analytic form for the pair-annihilation cross sections in Eqs. (\ref{eqannihilation}) - (\ref{equation:cascade}). The ``funnel'' at low values of $g^\prime$ corresponds to the resonant annihilation mode, while the large $g^\prime$ vertical asymptote to the $m_\chi\ll m_{Z^\prime}$ regime described above and, finally, the constant $g^{\prime2}/m_\chi$ region to the asymptotic  $m_\chi\gg m_{Z^\prime}$ regime (which onsets fully at larger $m_\chi$ than those in the figure).

\subsection{Indirect Detection}

Several probes exist of the late-time, low relative-velocity dark matter pair-annihilation cross section, generically known as ``indirect dark matter detection''. The characteristic feature of the model under consideration is the production of abundant hard leptons, as illustrated by the annihilation diagrams shown in Fig.~\ref{fig01}. As such, one possible constraint stems from cosmic-ray and gamma-ray data. In particular, the AMS-02 measurement of the cosmic ray positron fraction can be used to set constraints on dark matter annihilating to leptons \cite{Bergstrom:2013jra,Ibarra:2013zia,Kopp:2013eka,Lu:2015pta,DiMauro:2015jxa}. Dark matter below $\sim 100$~GeV annihilating into muons is in principle subject to such constraints. However, utilizing charged cosmic rays to set constraints on dark matter annihilation inevitably involves significant uncertainties from propagation and energy losses in the Galaxy, and  and, in the case of light dark matter, $m_\chi\lesssim 15$ GeV, also from solar modulation.  Though gamma-rays offer a promising search channel for dark matter annihilation, in the case of leptonic interactions, they are not the best probe. A more model-independent and robust way to probe the pair-annihilation rate is to use the effects of energy injection from dark matter annihilation on the cosmic microwave background (CMB). Specifically, dark matter annihilation at redshifts $z \sim 1000$ results in energy injection which heats and ionizes the photon-baryon plasma,  significantly perturbing the ionization history in a way that can be constrained by measurements  of the CMB temperature and polarization angular power spectra~\cite{Slatyer:2009yq}.  In our model, where the SM annihilation final states largely involve charged leptons, a sizable fraction, and in some case almost all of the energy is deposited in energetic electrons and positrons, the species with the highest effective deposited power fraction $f_{eff}$~\cite{Slatyer:2009yq}.

Constraints from CMB distortions are 
largely insensitive to systematic uncertainties \cite{Finkbeiner:2011dx,Galli:2011rz,Weniger:2013hja,Lopez-Honorez:2013lcm,Madhavacheril:2013cna,Galli:2013dna,Slatyer:2015kla}, and with the latest CMB data from Planck correspond to the limit~\cite{Ade:2015xua}
\begin{equation} \label{eq:CMB}
f_{eff} \frac{\sigma v}{ m_{\chi}} \lesssim 3 \times 10^{-28} {\rm cm^3/s/GeV} \,.
\end{equation}
In the equation above, $f_{eff}$ quantifies the efficiency with which the energy deposited per annihilation is actually injected at a given redshift in the universe history. 
This efficiency depends on the annihilation products.
For heavy $Z^\prime$ bosons, $m_{Z^\prime} > m_\chi$, our dark matter annihilates into muons, taus, and neutrinos. Neglecting phase space effects, we find the following approximate relative ratios 
\begin{equation}
 \mu^+\mu^- \Big/ \tau^+\tau^- \Big/ \nu_\mu \bar\nu_\mu + \nu_\tau \bar\nu_\tau  ~~\simeq~~  33.3\% \Big/ 33.3\% \Big/ 33.3 \% ~.
\end{equation}
If the $Z^\prime$ boson is sufficiently light also $\chi \bar\chi \to Z^\prime Z^\prime$ annihilations are possible. Neglecting again phase space effects and setting $q_\chi = q_\ell$, we find approximately
\begin{equation}
 \mu^+\mu^- \Big/ \tau^+\tau^- \Big/ \nu_\mu \bar\nu_\mu + \nu_\tau \bar\nu_\tau \Big/ Z^\prime Z^\prime ~~\simeq~~  25\% \Big/ 25\% \Big/ 25\% \Big/ 25\% ~,
\end{equation}
with the $Z^\prime$ bosons decaying back to neutrinos, muons and taus, if kinematically allowed.

In the case of 100\% annihilation into muons, for which $f_{eff} \simeq 0.2$, the latest results from Planck~\cite{Ade:2015xua} solidly exclude the canonical annihilation cross section of $3 \times 10^{-26} {\rm cm^3/s}$ for dark matter masses below $\sim 20$~GeV~\cite{Ade:2015xua}, when no velocity-dependence exists in the pair-annihilation cross section of the Dirac dark matter. Pair-annihilation to taus, which also includes hadronic final states, leads to similar results since it has only a slightly lower $f_{eff}$. 
Annihilation to four muons (which is relevant in the case of $\chi \bar\chi \to Z^\prime Z^\prime \to 4 \mu$) has also been shown to correspond to $f_{eff} \sim 0.2$~\cite{Slatyer:2015jla}, leading therefore to similar conclusions. Annihilations into neutrinos produce negligible effects to the CMB power spectrum. In the latter scenario, neutrino \cite{Adrian-Martinez:2016gti,Adrian-Martinez:2015wey} and gamma-ray detectors provide a better probe \cite{Queiroz:2016zwd}, but these are still far from the canonical cross section, thus placing no relevant constrain to our model.

The portion of the pair-annihilation cross section giving rise to the right amount of thermal relic dark matter (green curve) which is excluded by current Plack data is shaded in blue and labeled by {\it CMB} in Fig.~\ref{queirozplot01}.
In addition to the current Planck results, we will also quote a forecast for a cosmic-variance-limited experiment with similar angular resolution which might improve the current sensitivity by a factor of four \cite{Ade:2015xua}. The corresponding region is shaded in dark blue and labeled as {\it CMB proj.} in Figs. \ref{queirozplot01} - \ref{queirozplot7}.

The canonical annihilation cross section is excluded for $m_{\chi} \lesssim 20$~GeV for 100\% annihilation into muons using current CMB data. However, in our case, a sizeable fraction of the annihilation rate goes into neutrinos that yield negligible constraints. For this reason we can see that for $m_{\chi}=(10-15)$~GeV, Figs. \ref{queirozplot01}-\ref{queirozplot6}, only the projected CMB sensitivity applies.
It is important to emphasize that, at the resonance, the direct relation between the annihilation cross section today and at the time of freeze-out might fail, as pointed out in Ref.~\cite{Griest:1990kh,Edsjo:1997bg}, due to thermal effects in the early universe, explaining why the resonance region in the figures is not excluded. 

\subsection{Direct Detection}
\label{section: directdetection}

\begin{figure}[!t]
\centering
\includegraphics[scale=0.6]{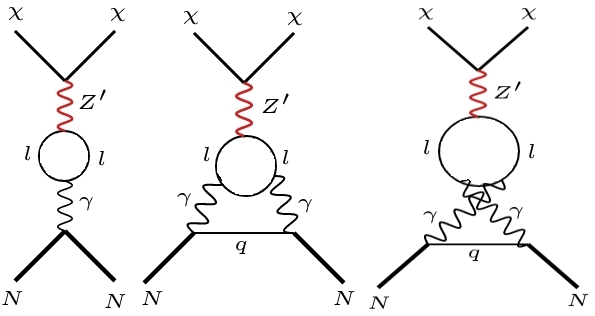}
\caption{Diagrams producing spin-independent dark matter-nucleon scattering at the loop level.}
\label{figure:DD}
\end{figure}

Since the DM particle does not directly couple to quarks, scattering off of nuclei occurs only through a loop of charged leptons that couple to photons which in turn couple to protons, as illustrated in Fig.~\ref{figure:DD}~\footnote{Strictly speaking, in order to explain the $B$ decay anomalies, the $Z^{\prime}$ boson has to couple to SM quarks. In~\cite{Altmannshofer:2014cfa} this is achieved through dimension-six effective operators or through heavy vector-like quarks, which mix with SM ones, inducing a small flavor violating $Z^{\prime}bs$ coupling. While a flavor-violating coupling cannot lead to any appreciable contribution to the direct detection cross section, it might be accompanied by flavor diagonal couplings to quarks. The size of the flavor diagonal couplings is model dependent and we will not consider their effects here.}.
The WIMP-nucleon scattering cross section is thus proportional to the nucleus electric charge. Moreover, since the $Z^{\prime}$ boson has vector-like interactions with the charged leptons and the dark matter, the scattering cross section is not velocity dependent.
The left diagram in Fig.~\ref{figure:DD} is the most relevant, while the other two are 2-loop suppressed and therefore negligible. In our numerical calculations we include only the former.
Adapting the results obtained in \cite{Kopp:2009et} for leptophilic dark matter to our model, we find the WIMP-nucleon scattering cross section to be 
\begin{equation}\label{eq:dirdet}
\sigma_{SI} =\frac{1}{A^2} \frac{\mu_{N}^2}{9\pi} \left(\frac{\alpha_\text{em}\, Z\, g^{\prime 2} q_{\chi} q_l }{\pi m_{Z^{\prime}}^2}\log \left( \frac{m_{\mu}^2}{m_{\tau}^2} \right)\right)^2 \,,
\end{equation}
where $\mu_N= m_N m_{\chi}/(m_N+m_{\chi})$ is the WIMP-nucleus reduced mass, $m_N$ the nucleus mass, and $Z$ and $A$ the atomic and mass numbers, respectively. 
Since we will compare our theoretical predictions with current limits from Xe-based experiments, $Z=54$ and $m_N \simeq 129$~GeV. The dependence on the logarithm of the lepton masses squared can be understood as a leading log approximation to the RGE induced kinetic mixing between the $Z^\prime$ and the photon in the running from the tau mass down to the muon mass.

\begin{figure}
\centering
\includegraphics[scale=0.61]{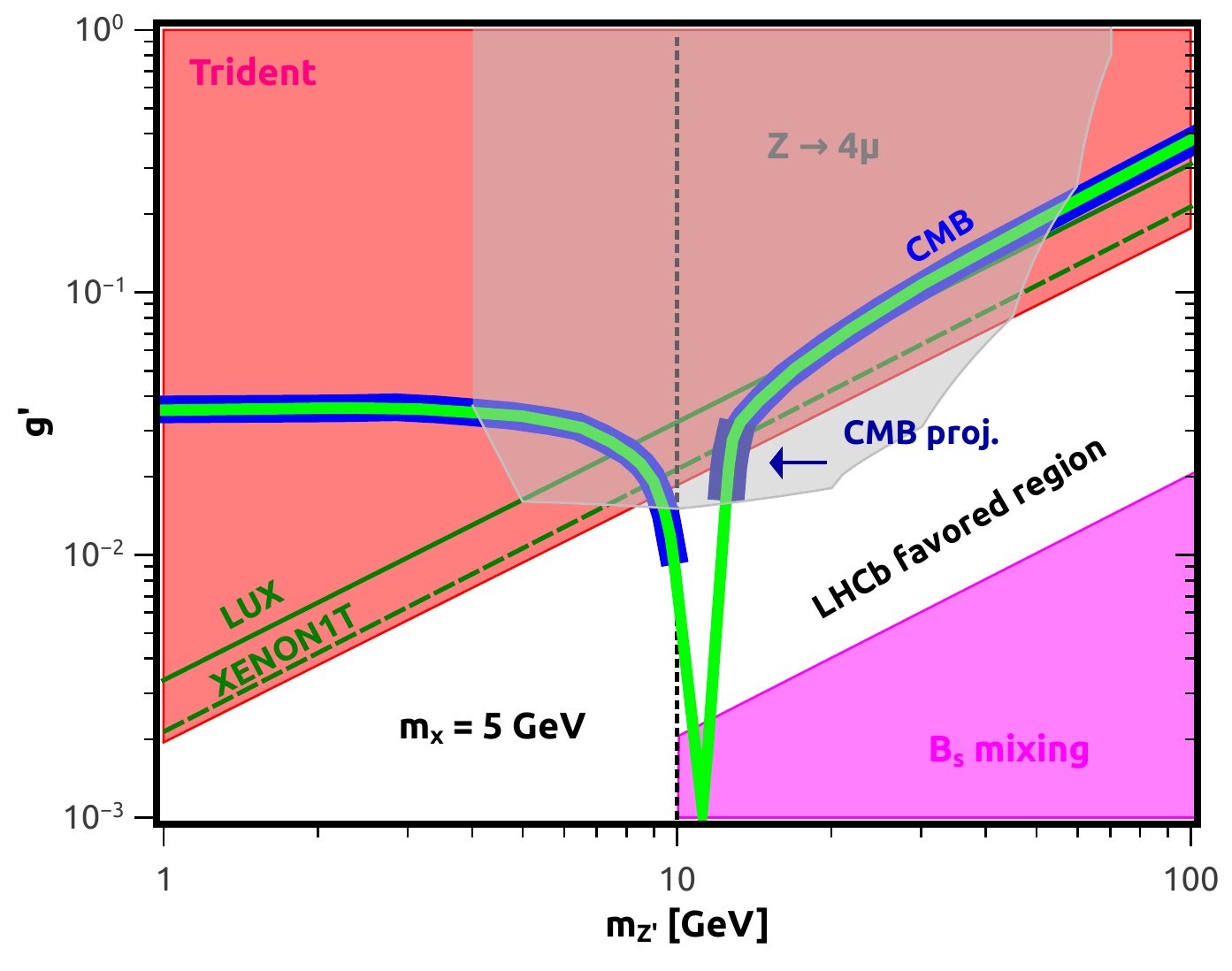}\\
\includegraphics[scale=0.61]{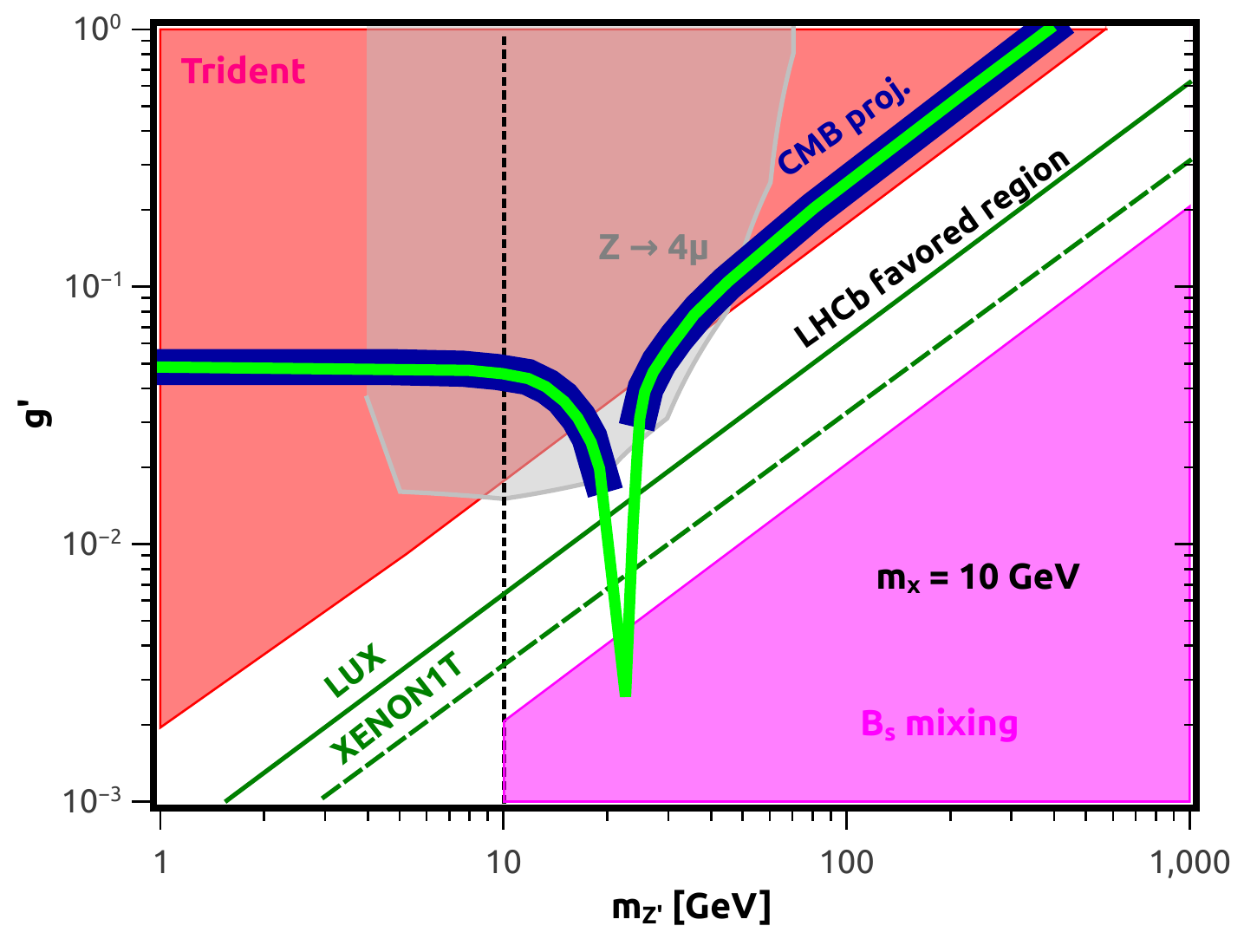}
\caption{The $(m_{Z^\prime}, g^\prime)$ parameter space at fixed values for the dark matter mass $m_\chi=$ 5~GeV (upper panel) and 10~GeV (lower panel). The charge of the DM is always fixed to $q_\chi = 1$. The region able to explain the rare $B$ decay anomalies, while being not ruled out by other experiments, is the central white diagonal band, for $m_{Z^\prime} \gtrsim 10$~GeV. The light green curve corresponds to the parameters producing the right thermal relic dark matter abundance, with the portions shaded in blue excluded by CMB data at present (light blue) or in the future (darker blue). The shaded magenta and red regions are excluded by $B_s$ mixing and neutrino trident production observables, respectively. The gray contour delimits the region ruled out by the measurement of the $Z \rightarrow 4\mu$ decay width. The dark green solid diagonal line represents the LUX-2016 direct dark matter detection bound, with the parameter space to the upper left being excluded.
The dashed dark green diagonal line corresponds to the expected XENON1T sensitivity with 2 year data.}
\label{queirozplot01}
\end{figure}

\begin{figure}
\centering
\includegraphics[scale=0.61]{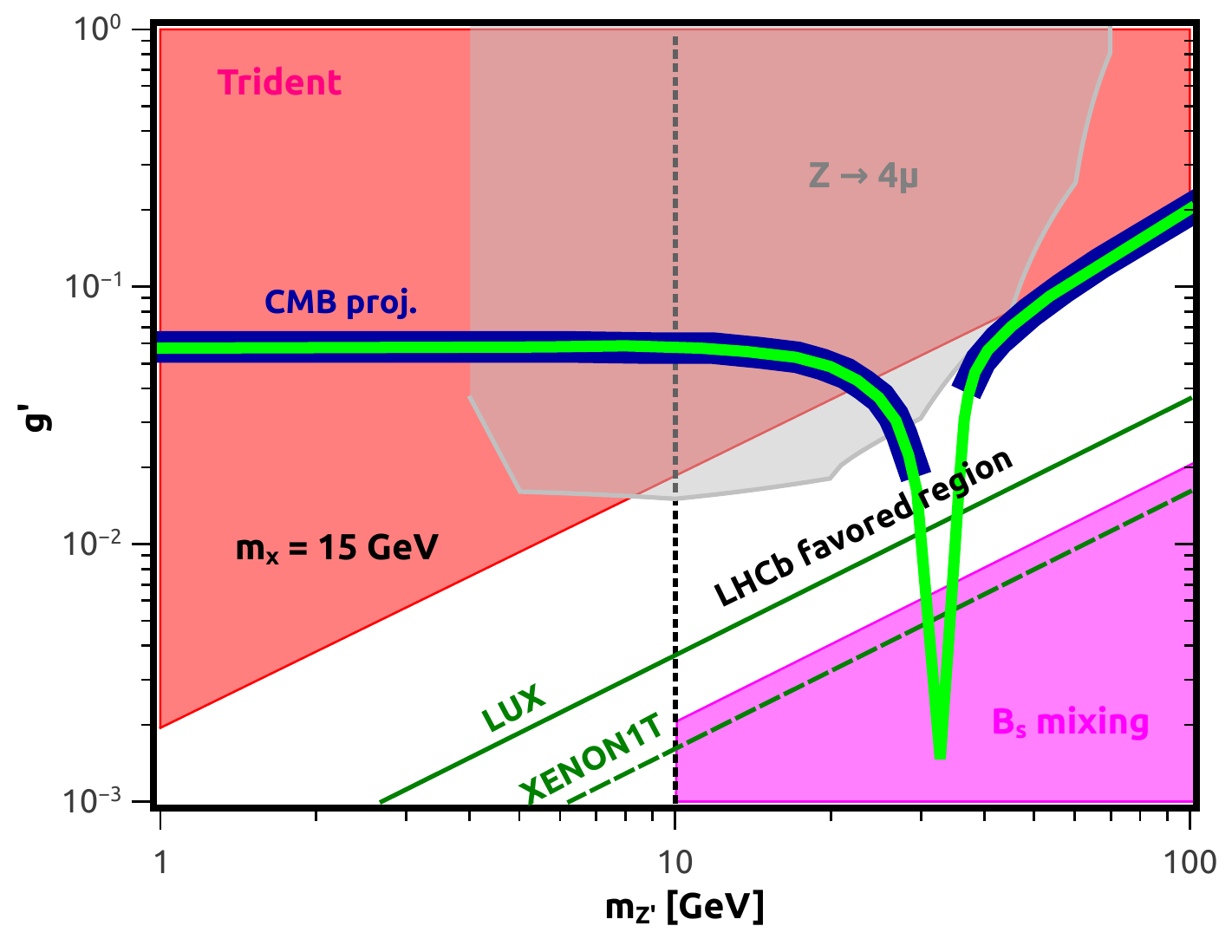}\\
\includegraphics[scale=0.61]{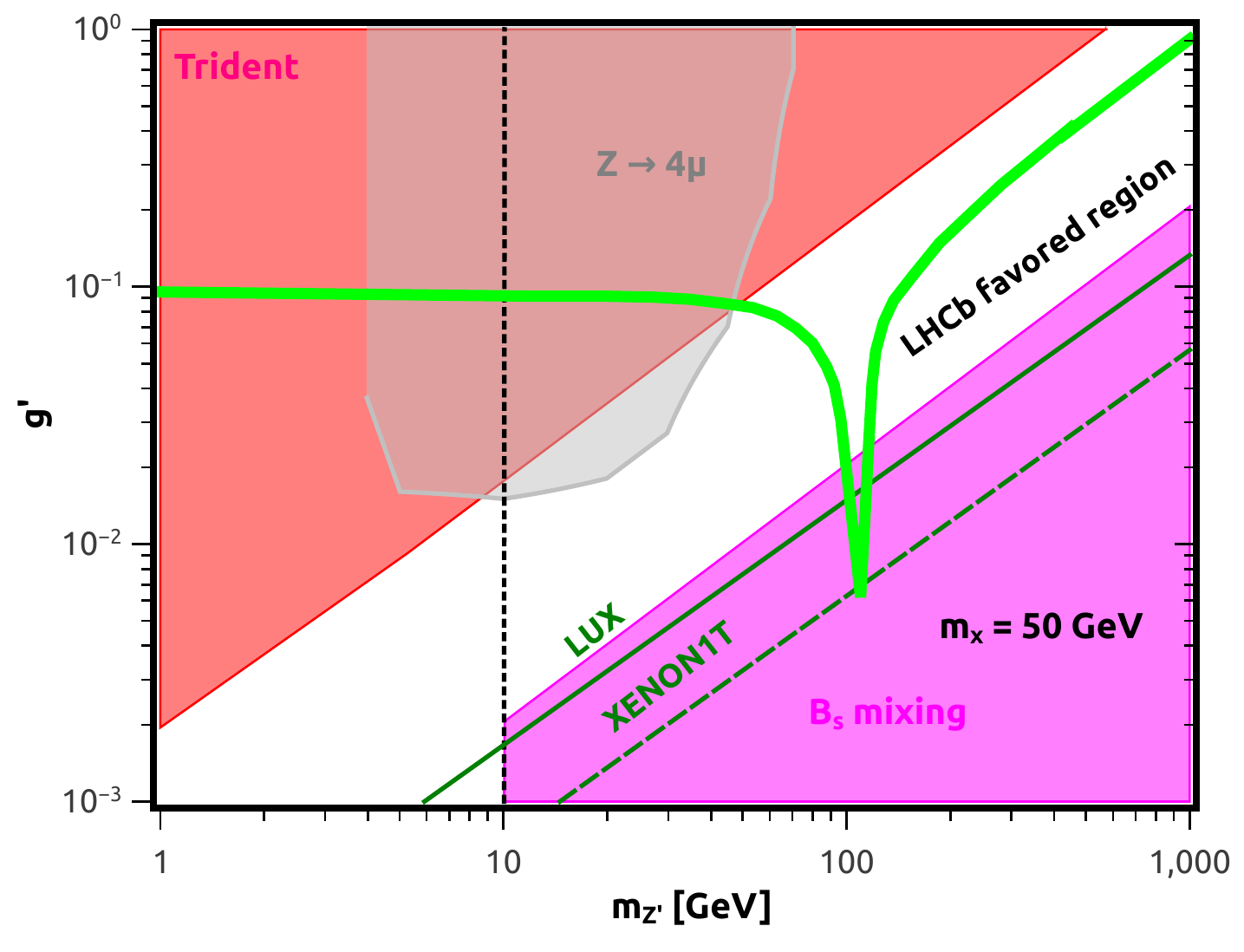}\\
\includegraphics[scale=0.61]{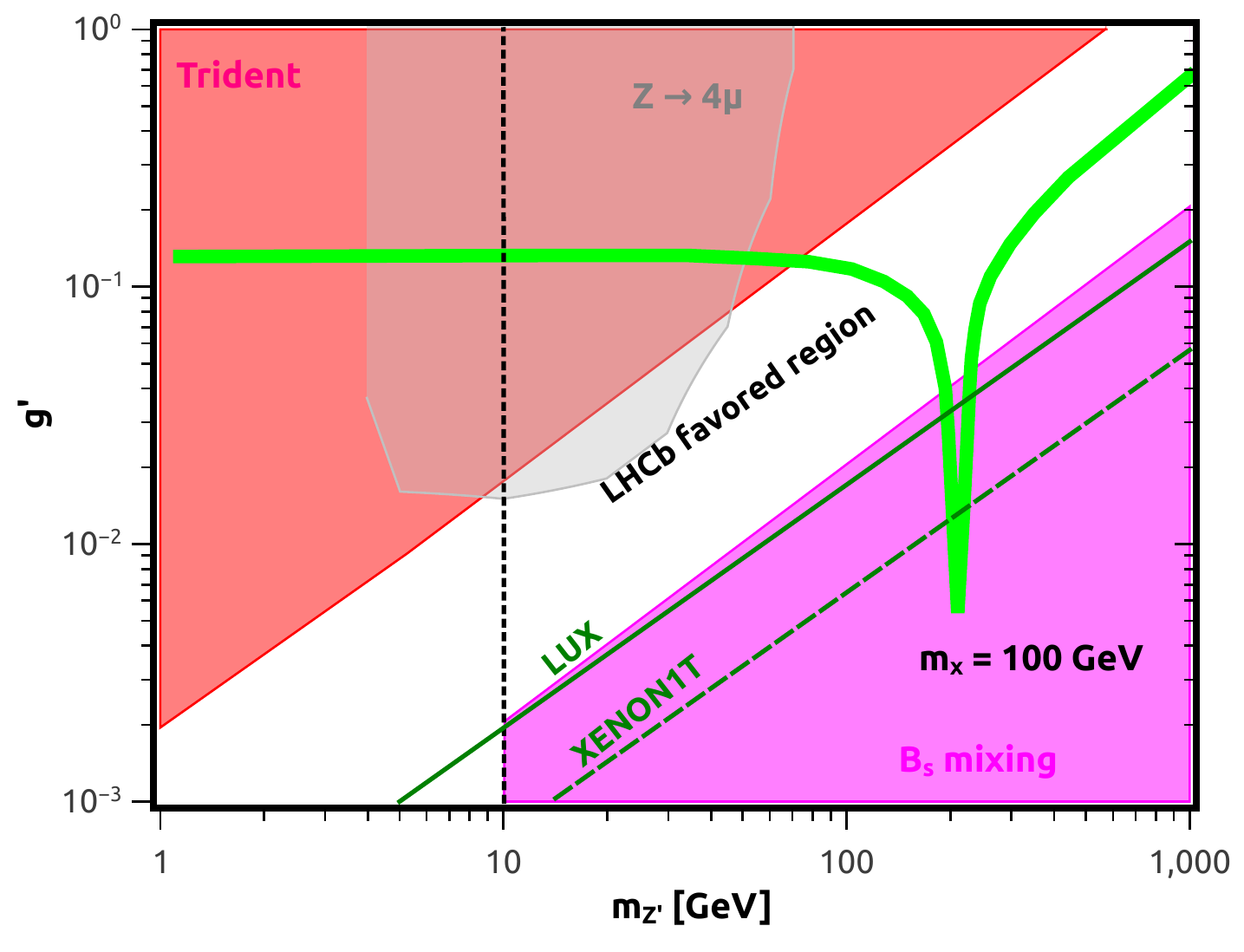}
\caption{
As in Fig.~\ref{queirozplot01} but for dark matter masses $m_\chi=$ 15~GeV (top panel), 50~GeV (center panel) and 100~GeV (bottom panel); $q_\chi=1$.
}
\label{queirozplot1}
\end{figure}

\begin{figure}
\centering
\includegraphics[scale=0.61]{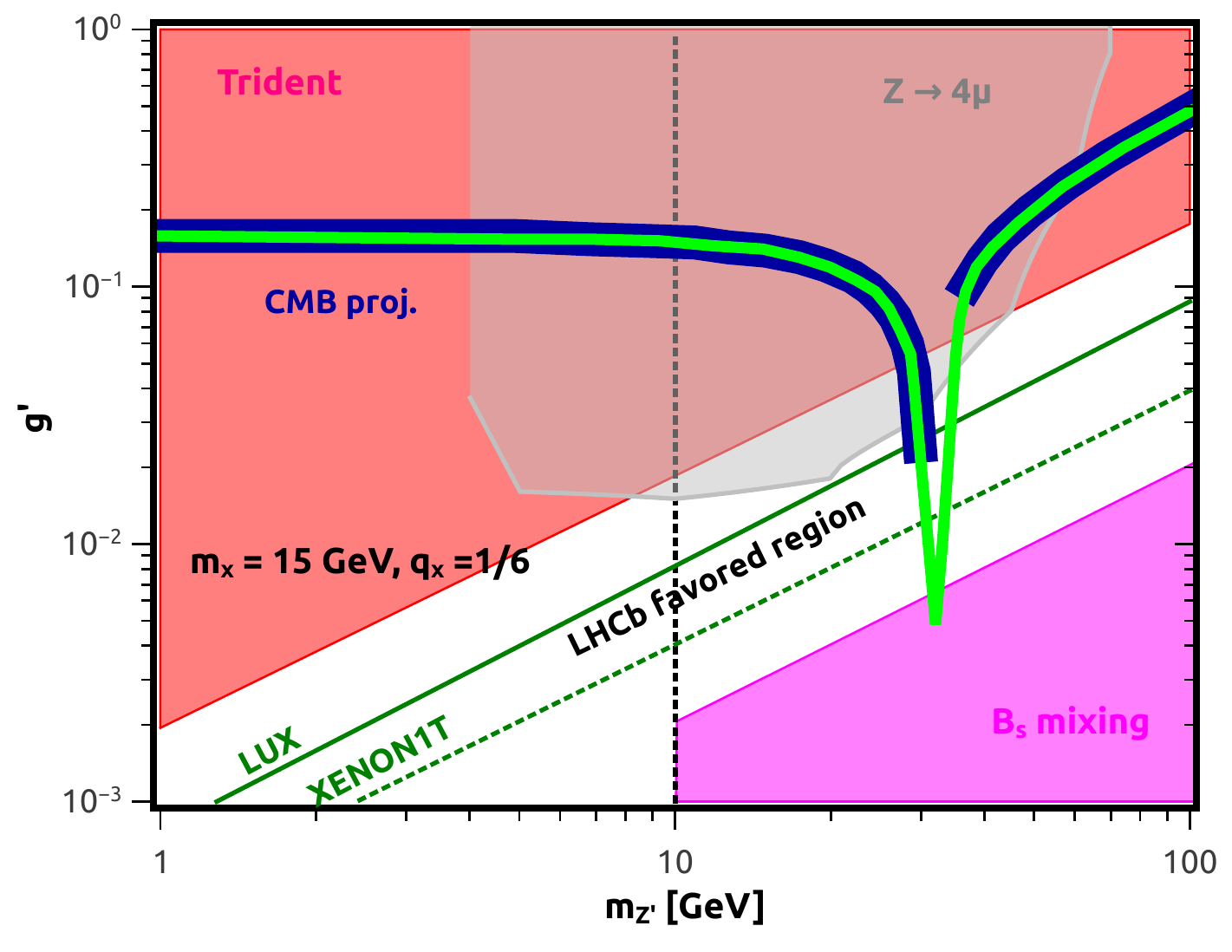}\\
\includegraphics[scale=0.61]{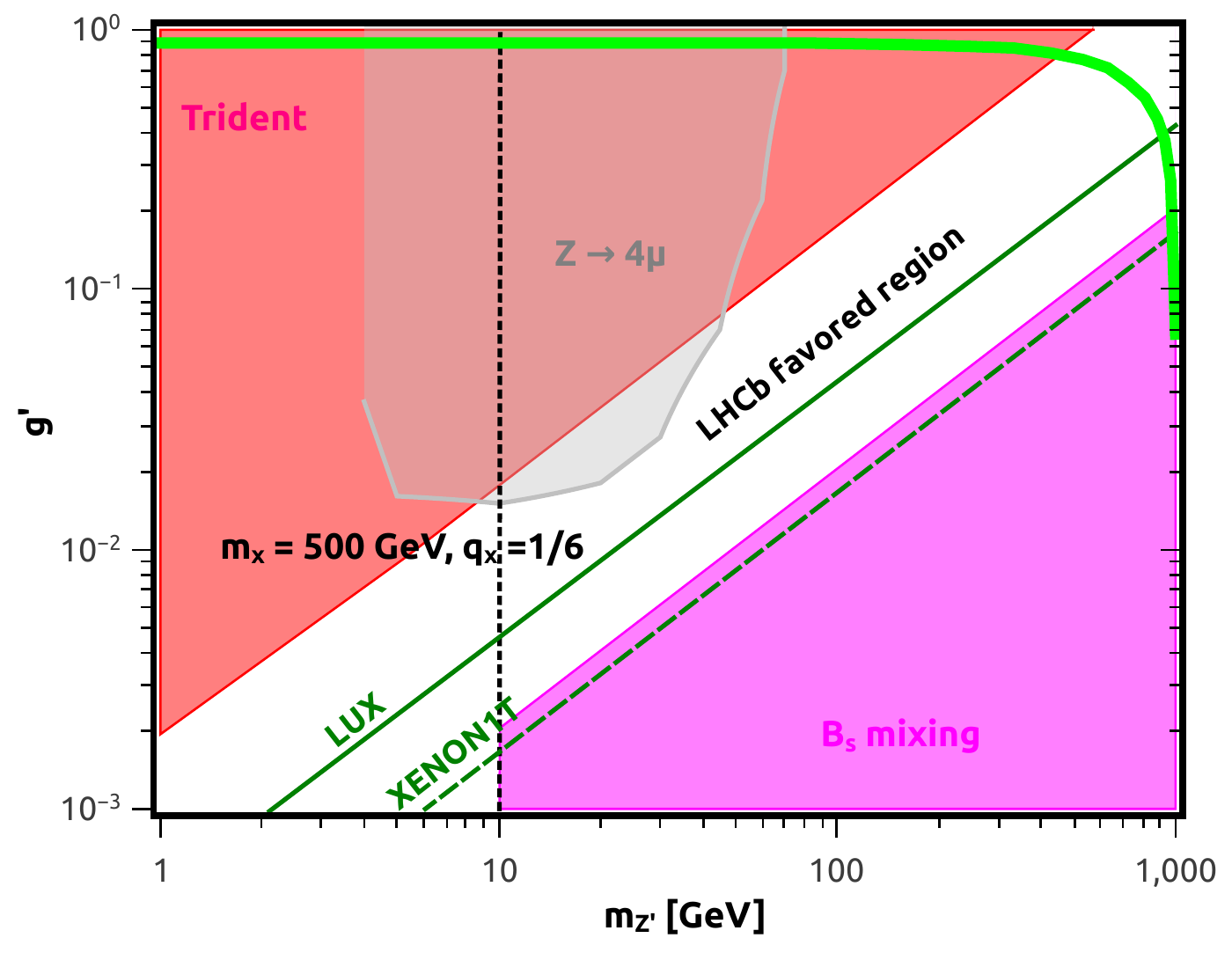}
\includegraphics[scale=0.61]{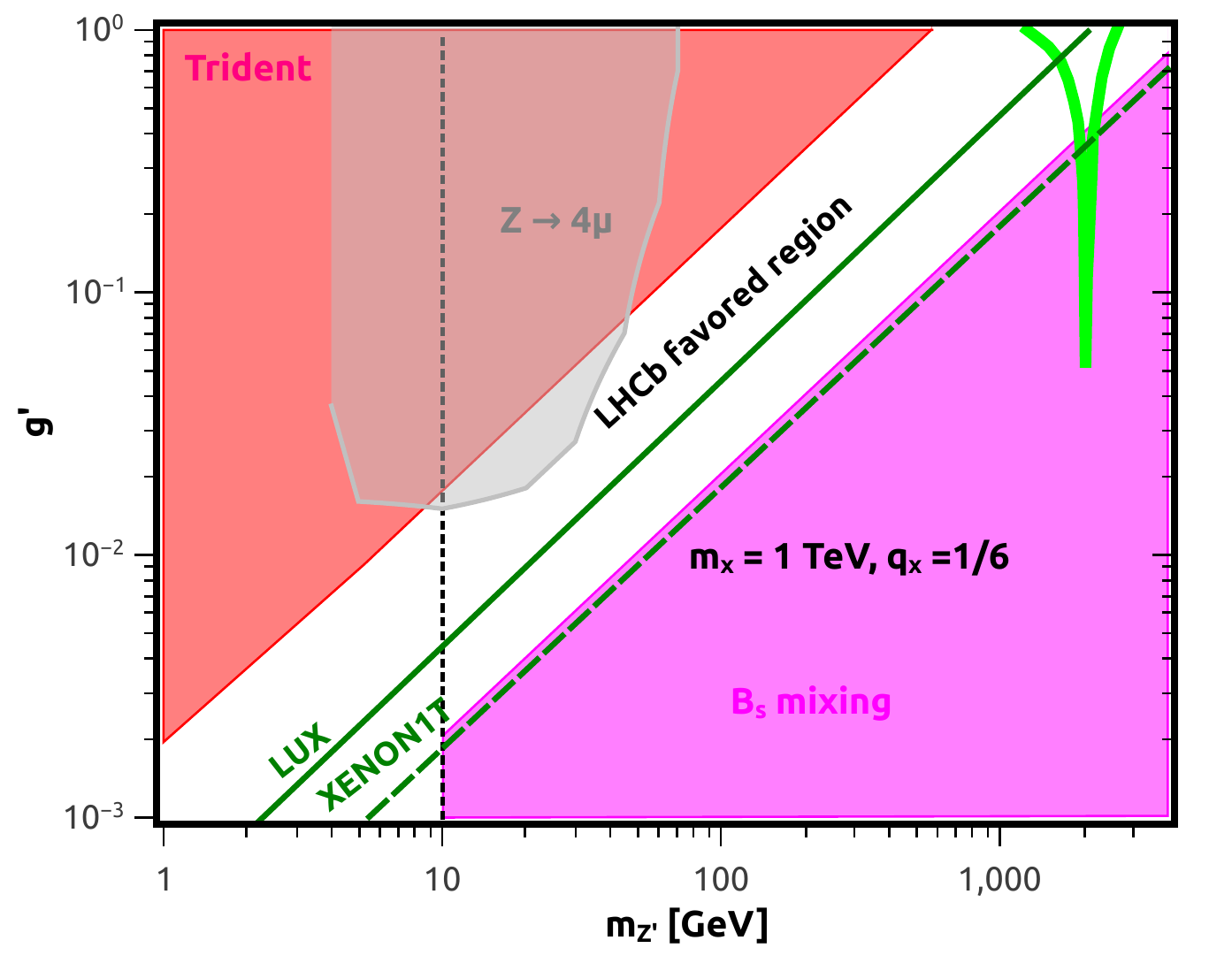}
\caption{
As in Fig.~\ref{queirozplot01}, but for $q_{\chi}=1/6$ and for the dark matter masses $m_\chi=15$~GeV (top panel), 500~GeV (center panel) and $1$~TeV (bottom panel).
}
\label{queirozplot6}
\end{figure}

\begin{figure}
\centering
\includegraphics[scale=0.61]{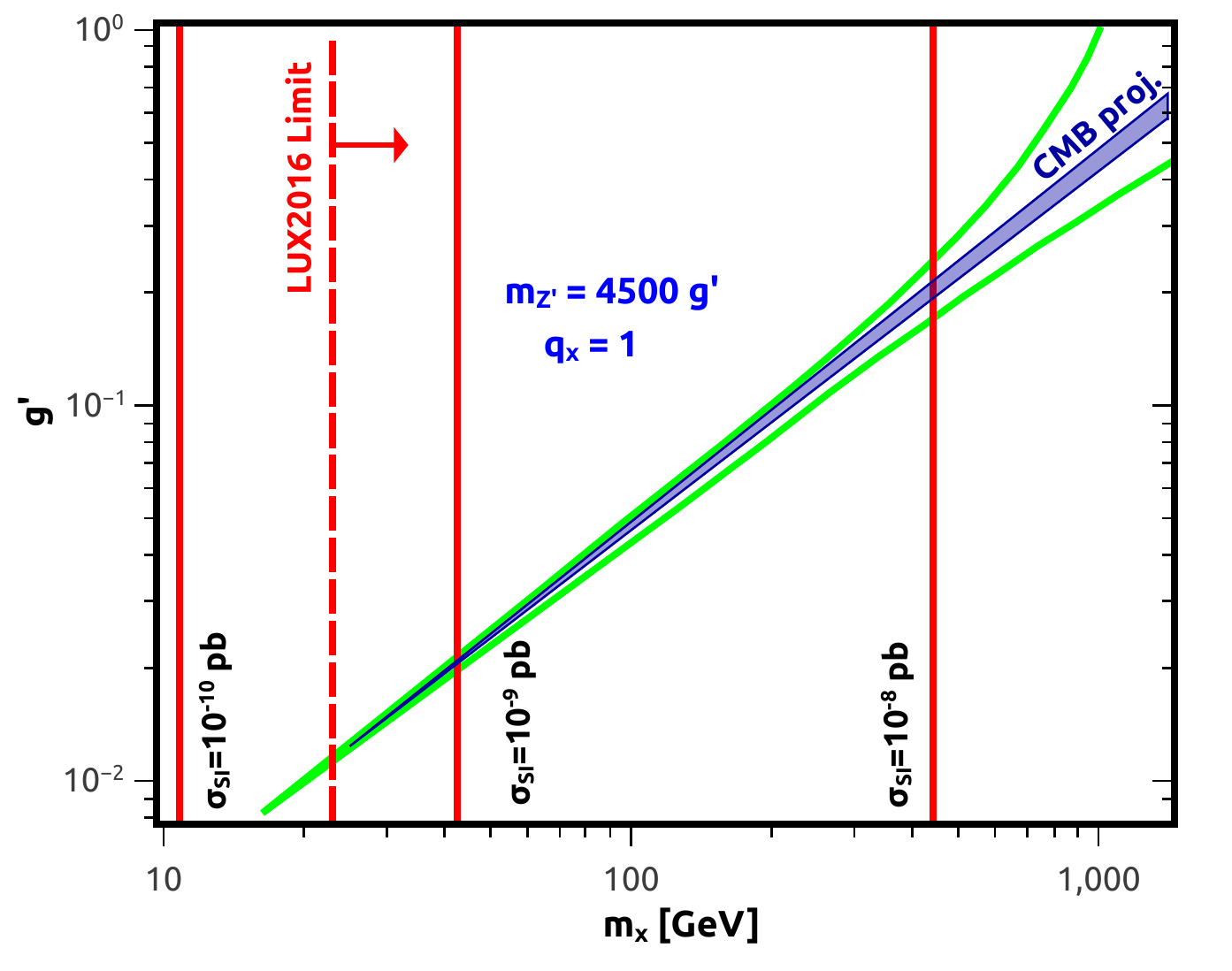}\\
\includegraphics[scale=0.61]{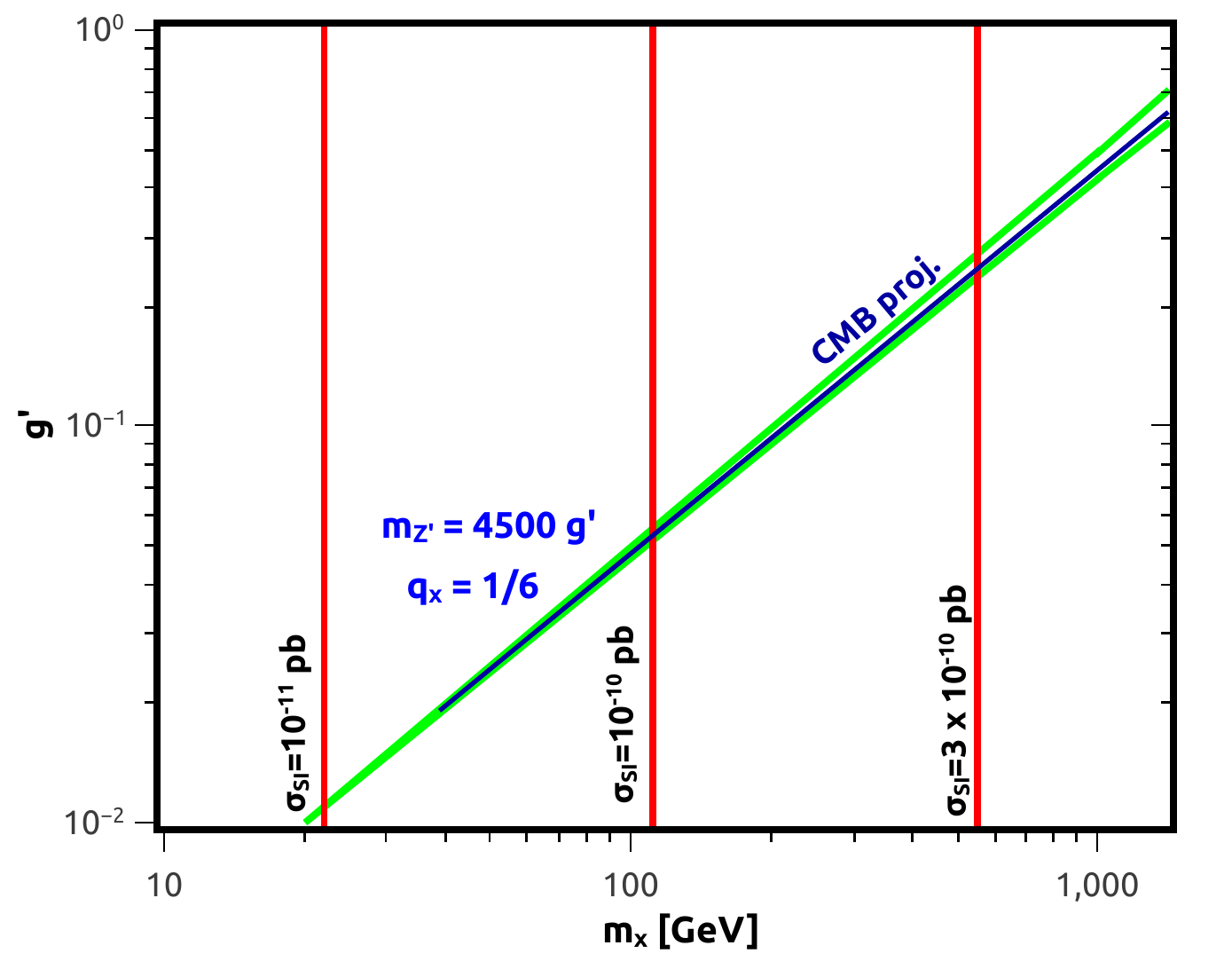}
\caption{The $(m_\chi,g^\prime)$ parameter space. The mass of the $Z^\prime$ is set to a benchmark value to explain the rare $B$ decay anomalies $m_{Z^\prime}=g^\prime \times 4500$~GeV, with $q_{\chi}=1$ (upper panel), and $q_{\chi}=1/6$ (lower panel). The green region corresponds to the parameter space reproducing the right thermal relic abundance. The red vertical lines indicate selected values for the spin-independent direct detection cross section. The blue shaded region can be probed by projected sensitivities to CMB distortions. In the upper panel, the dashed red vertical line indicates the present LUX bound.}
\label{queirozplot7}
\end{figure}

The most stringent limits on the WIMP-nucleon spin-independent scattering cross section stem at present from the LUX experiment~\cite{Akerib:2016vxi}, which exclude $\sigma_{SI}  \gtrsim 2 \times 10^{-10} {\rm pb}$  for $m_{\chi} \sim 50$~GeV, surpassing  earlier results~\cite{Akerib:2013tjd,Akerib:2015rjg} after achieving better calibration, event-reconstruction, background modeling, and more live-days. Similar sensitivity has been achieved by PANDAX-II \cite{Tan:2016zwf}. We present the current LUX and the projected Xenon1T sensitivities in solid green in Figs.~\ref{queirozplot01}-\ref{queirozplot6}, for a variety of choices for the dark matter particle mass, set to 5, 10 (Fig.~\ref{queirozplot01}), 15, 50 and 100 GeV  (Fig.~\ref{queirozplot1}). The figures also show the projected sensitivity from the XENON1T experiment with 2 years of data~\cite{Aprile:2015uzo}, expected to improve the current LUX limits by two orders of magnitude (dashed green line).
To obtain these curves we assume canonical values for the local dark matter density and velocity distribution throughout the plots.

The figures illustrate that direct detection provides remarkably strong limits in spite of being loop-suppressed in our model. Direct detection constraints exclude dark matter masses above $\sim 15 $~GeV, allowing only the peak corresponding to the $Z^{\prime}$ resonance. For $m_{\chi} < 15$~GeV, direct detection limits weaken, because for lower masses the recoil energy is sufficiently small and close to the energy threshold of the experiments, degrading the corresponding sensitivity.
This conclusion depends on the fact that we have fixed $q_\ell=q_\chi$. In the next section we will also depart from this equality and address the consequences. The corresponding results are collected in Fig. \ref{queirozplot6}.

\section{The Global Picture: Dark Matter and $B$ Anomalies}\label{sec:global}

In this section we combine all findings described in the previous sections and outline the region of parameter space which can simultaneously accommodate thermal relic dark matter and the rare $B$ decay anomalies. The region in the $(m_{Z^\prime}, g^\prime)$ parameter space that accommodates the $B$ physics anomalies corresponds to the central diagonal white regions in Figs.~\ref{queirozplot01}-\ref{queirozplot6}. The top red and bottom magenta shaded regions are ruled out by neutrino trident production and $B_s$ mixing, respectively. The gray contour delimits the region ruled out by the measurement of the $Z \rightarrow 4\mu$ decay width. Note that the region favored by the flavor anomalies ends at $\sim 10$~GeV, as discussed in Sec.~\ref{sec:anomalies}. The light green curves indicate $(m_{Z^\prime}, g^\prime)$ combinations that reproduce the right relic abundance ($\Omega_{\chi} h^2 \simeq 0.12$). Relic abundance curves with the light (dark) blue contours are excluded by  limits from current (projected) CMB data. Diagonal solid (dashed) green lines indicate the LUX (the projected XENON1T) bounds: the region to the top left of those lines is excluded.

In Fig.~\ref{queirozplot01} we show results for $m_{\chi}=5, 10$~GeV and charge $q_\chi=1$. For $m_{\chi}=5$~GeV, current and projected direct detection limits (green curves on the top left of the plots) are weak due to the experimental energy threshold. The correct relic abundance can be achieved close to the resonance region, without being excluded by the constraints from the neutrino trident production.
The CMB limit probes a sizable part of parameter space that yields the right abundance. Forecasted CMB data constrain the model a bit further. In the $m_{\chi}=5$~GeV case we are left with a small region of parameter space with a $Z^\prime$ mass of $m_{Z^\prime} \simeq 10$~GeV and a $L_\mu - L_\tau$ gauge coupling of $g^\prime \sim 10^{-3} - 10^{-2}$.

For $m_{\chi}=10$~GeV, only the projected CMB sensitivity applies. Direct detection limits from LUX and projected limits from XENON1T probe substantial parts of the parameter space of the model where one can accommodate both the LHCb anomaly and the relic density. Notice that CMB limits weaken at the resonance because, at the resonance, sufficiently small couplings can still reproduce the right abundance, however such small coupling lead to small annihilation cross sections today. This trend can be seen for $m_{\chi}=(5-15)$~GeV (see the upper panel of Fig. \ref{queirozplot1} for the latter case).  

Fig.~\ref{queirozplot1} focuses on dark matter masses $m_{\chi} =15,\ 50, \ 100$~GeV. Notice that direct detection limits (green curves on the top left of the plots) are significantly stronger than in the $5$~GeV case, becoming most relevant for a $50$~GeV dark matter mass; XENON1T projected sensitivity can entirely probe the LHCb favored region for $m_\chi = 15$~GeV. The entire favored parameter space for $m_\chi = 50$~GeV and for $m_\chi = 100$~GeV is already excluded by current LUX bounds combined with $B_s$ mixing.
We find that the current LUX bounds leaves viable parameter space only for $m_\chi \lesssim 23$~GeV.

In summary, in the regime with $q_l \simeq q_\chi$, the dark matter mass window of $m_\chi \simeq (5 - 23)$~GeV is favored. For smaller dark matter masses the $Z^\prime$ required to accommodate the right relic abundance is too light to explain the $B$ decay anomalies. For larger dark matter masses direct detection and $B_s$ mixing exclude the entire favored parameter space.
Upcoming XENON1T data might significantly test the model, leaving only masses around $m_\chi \simeq (5 - 10)$~GeV open. The latter can be later probed with further improvements from $B_s$ mixing constraints, and the LZ direct detection experiment~\cite{Akerib:2015cja}. 

These conclusions change if we depart from the assumption that $q_l \simeq q_\chi$. In Fig.~\ref{queirozplot6} we show the results for $q_{\chi}=1/6$, keeping $q_l=1$. The indirect detection limits are essentially the same since the smaller $q_{\chi}$ the larger the gauge coupling needed to obtain the right relic abundance, yielding no impact in the overall annihilation cross section needed to get the relic density. On the order hand, direct detection limits are weaker ameliorating the scenario specially for heavy masses. Indeed, for the examples $m_{\chi}=500$~GeV and 1~TeV, one can see viable regions of parameter space in the center and bottom panels of Fig.~\ref{queirozplot6} that are able to simultaneously accommodate dark matter and the LHCb anomaly.

Our final Fig.~\ref{queirozplot7} provides an orthogonal view of the parameter space in the $(m_\chi,g^\prime)$ plane, with the mass of the $Z^\prime$ set to $m_{Z^\prime}=g^\prime \times 4500$~GeV which is inside the LHCb favored region for both choices of $q_\chi$:  $q_\chi=1$ (upper panel) and $q_\chi=1/6$ (lower panel). In this regime, CMB limits do not touch the relic density curves (green) but do probe the funnel region with larger annihilation cross sections. In the figures we show these limits in dark blue. For $q_{\chi}=1/6$, the funnel region is rather narrow and for this reason the CMB projected limit is simply a line. The vertical red lines show contours of constant spin-independent dark matter-nucleon cross section $\sigma_{SI}$. Notice that since $\sigma_{SI}\propto g^{\prime4}/m_{Z^\prime}^4$ and the ratio $g^\prime/m_{Z^\prime}$ is kept fixed in the figure, the cross section only depends on the dark matter mass through the reduced mass, $\mu_N$ (see Eq.~(\ref{eq:dirdet})). Having in mind that current LUX limit excludes $\sigma_{SI} \simeq 4 \times 10^{-9}$~pb for $m_{\chi}=10$~GeV and $\sigma_{SI} \simeq 2.5 \times 10^{-10}$~pb for $m_{\chi}=100$~GeV, we conclude that only light dark matter can circumvent all existing constraints and accommodate the $B$ decay anomalies if $q_{\chi}=1, q_{\ell}=1$ (upper panel of Fig. \ref{queirozplot7}). For the $Z^\prime$ mass of $m_{Z^\prime}=g^\prime \times 4500$~GeV we find the bound $m_\chi \lesssim 22.9$~GeV as indicated by the dashed vertical line in the plot.  

In the bottom panel, we set $q_{\chi}=1/6,q_{\ell}=1$, weakening direct detection limits. In this case, the entire parameter space is consistent with the data and can address the $B$ decay anomalies.

\section{Conclusions}\label{sec:conclusions}

In this paper we proposed a new physics setup with dark matter charged under a $L_\mu-L_\tau$ gauge symmetry. Specifically, we studied if one could address at the same time anomalies observed in rare $B$ meson decays by the LHCb collaboration and have a viable thermal dark matter candidate. Our model hinges upon the (automatically anomaly-free) $L_{\mu}-L_{\tau}$ local gauge symmetry, $U(1)_{\mu-\tau}$, with a new corresponding massive gauge boson $Z^\prime$. Dark matter is a vector-like Dirac fermion charged under the $L_\mu - L_\tau$ symmetry and neutral under the SM gauge symmetries. This setup leads to an unusual and novel dark matter phenomenology, which we described in detail.

The dark matter relic abundance is mainly set by annihilation into muon, tau, and neutrino pairs, through $s$-channel $Z^\prime$ exchange.
We studied indirect detection limits stemming from distortions of the CMB power spectrum, and direct detection using current LUX2016 and XENON1T projected sensitivities. Despite the fact that dark matter nucleus scattering is loop-suppressed, we found that direct detection experiments do put remarkably strong constraints on the model parameter space.
The correct relic abundance can only be obtained close to the resonance region $m_{Z^\prime} \simeq 2 m_\chi$. 
Combining the direct detection constraints with constraints from $B_s$ meson oscillations leaves only a restricted window of parameter space where both $B$ decay anomalies and the correct relic abundance can be explained: the dark matter mass is $5~\text{GeV} \lesssim m_\chi \lesssim 23$~GeV; the $Z^\prime$ mass is $m_{Z^\prime} \simeq 2 m_\chi$; the gauge coupling is in the range $2 \times 10^{-3} \lesssim g^\prime \lesssim 10^{-2}$.
Substantial parts of this parameter space can be probed by the expected sensitivities of future direct detection experiments like XENON1T and LZ.

The viable parameter space can be extended significantly by allowing the dark matter to have $L_\mu - L_\tau$ charges that are smaller than the charges of the SM leptons. In this case direct detection constraints are weakened, and their upper bound on the dark matter mass disappears. A correct relic abundance still requires a $Z^\prime$ mass that is close to the resonance region. Prospects for probing such a scenario with future direct detection experiments are excellent.

\section*{Acknowledgements}

WA thanks Joachim Brod for useful discussions. WA and SG acknowledge support from the University of Cincinnati. WA and SG thank Dan Hooper for discussions about the AMS-02 bounds. WA and FSQ are grateful to the Mainz Institute for Theoretical Physics (MITP) for its hospitality and partial support during the final stages of this work. WA and SG thank the Aspen Center for Physics for hospitality during the completion of this work. The Aspen Center for Physics is supported by National Science Foundation grant PHY-1066293. SP is partly supported by the U.S. Department of Energy grant number DE-SC0010107.

\bibliographystyle{JHEPfixed}
\bibliography{darkmatter}

\providecommand{\href}[2]{#2}\begingroup\raggedright\begin{thebibliography}{100}

\bibitem{Aaboud:2016qgg}
{\bf ATLAS} Collaboration, M.~Aaboud {\em et.~al.}, {\it {Search for dark
  matter produced in association with a hadronically decaying vector boson in
  pp collisions at $\sqrt s $=13 TeV with the ATLAS detector}},
  \href{http://xxx.lanl.gov/abs/1608.02372}{{\tt 1608.02372}}.

\bibitem{ATLAS:2016tsc}
{\bf ATLAS} Collaboration, {\it {Search for Dark Matter production associated
  with bottom quarks with 13.3 fb$^{-1}$ of pp collisions at $\sqrt s$ = 13 TeV
  with the ATLAS detector at the LHC}},  2016.
\newblock ATLAS-CONF-2016-086.

\bibitem{Aaboud:2016tnv}
{\bf ATLAS} Collaboration, M.~Aaboud {\em et.~al.}, {\it {Search for new
  phenomena in final states with an energetic jet and large missing transverse
  momentum in $pp$ collisions at $\sqrt{s}=13$ TeV using the ATLAS detector}},
  {\em Phys. Rev.} {\bf D94} (2016), no.~3 032005,
  [\href{http://xxx.lanl.gov/abs/1604.07773}{{\tt 1604.07773}}].

\bibitem{Aaboud:2016uro}
{\bf ATLAS} Collaboration, M.~Aaboud {\em et.~al.}, {\it {Search for new
  phenomena in events with a photon and missing transverse momentum in $pp$
  collisions at $\sqrt{s}=13$ TeV with the ATLAS detector}},  {\em JHEP} {\bf
  06} (2016) 059, [\href{http://xxx.lanl.gov/abs/1604.01306}{{\tt
  1604.01306}}].

\bibitem{monoHiggsATLAS}
{\bf ATLAS} Collaboration, {\it {Search for Dark Matter in association with a
  Higgs boson decaying to $b$-quarks in $pp$ collisions at $\sqrt{s} =13$ TeV
  with the ATLAS detector}},  2016.
\newblock ATLAS-CONF-2016-019.

\bibitem{CMS:2016flr}
{\bf CMS} Collaboration, {\it {Search for new physics in a boosted hadronic
  monotop final state using $12.9~\mathrm{fb}^{-1}$ of
  $\sqrt{s}=13~\mathrm{TeV}$ data}},  2016.
\newblock CMS-PAS-EXO-16-040.

\bibitem{CMS:2016fnh}
{\bf CMS} Collaboration, {\it {Search for dark matter and graviton produced in
  association with a photon in pp collisions at $\sqrt s$ = 13 TeV}},  2016.
\newblock CMS-PAS-EXO-16-039.

\bibitem{CMS:2016hmx}
{\bf CMS} Collaboration, {\it {Search for dark matter in
  $\mathrm{Z}+E_\mathrm{T}^\mathrm{miss}$ events using $12.9~\mathrm{fb}^{-1}$
  of 2016 data}},  2016.
\newblock CMS-PAS-EXO-16-038.

\bibitem{CMS:2016pod}
{\bf CMS} Collaboration, {\it {Search for dark matter in final states with an
  energetic jet, or a hadronically decaying W or Z boson using
  $12.9~\mathrm{fb}^{-1}$ of data at $\sqrt{s} = 13~\mathrm{TeV}$}},  2016.
\newblock CMS-PAS-EXO-16-037.

\bibitem{CMS:2016mjh}
{\bf CMS} Collaboration, {\it {Search for dark matter in association with a
  Higgs boson decaying into a pair of bottom quarks at $\sqrt s$ = 13 TeV with
  the CMS detector}},  2016.
\newblock CMS-PAS-EXO-16-012.

\bibitem{CMS:2016xok}
{\bf CMS} Collaboration, {\it {Search for Dark Matter Produced in Association
  with a Higgs Boson Decaying to Two Photons}},  2016.
\newblock CMS-PAS-EXO-16-011.

\bibitem{CMS:2016mxc}
{\bf CMS} Collaboration, {\it {Search for dark matter in association with a top
  quark pair at $\sqrt s$ = 13 TeV}},  2016.
\newblock CMS-PAS-EXO-16-005.

\bibitem{CMS:2016uxr}
{\bf CMS} Collaboration, {\it {Search for Dark Matter produced in association
  with bottom quarks}},  2016.
\newblock CMS-PAS-B2G-15-007.

\bibitem{Altmannshofer:2014cfa}
W.~Altmannshofer, S.~Gori, M.~Pospelov, and I.~Yavin, {\it {Quark flavor
  transitions in $L_\mu-L_\tau$ models}},  {\em Phys. Rev.} {\bf D89} (2014)
  095033, [\href{http://xxx.lanl.gov/abs/1403.1269}{{\tt 1403.1269}}].

\bibitem{Altmannshofer:2015mqa}
W.~Altmannshofer and I.~Yavin, {\it {Predictions for lepton flavor universality
  violation in rare B decays in models with gauged $L_\mu - L_\tau$}},  {\em
  Phys. Rev.} {\bf D92} (2015), no.~7 075022,
  [\href{http://xxx.lanl.gov/abs/1508.07009}{{\tt 1508.07009}}].

\bibitem{He:1990pn}
X.~G. He, G.~C. Joshi, H.~Lew, and R.~R. Volkas, {\it {NEW Z-prime
  PHENOMENOLOGY}},  {\em Phys. Rev.} {\bf D43} (1991) 22--24.

\bibitem{He:1991qd}
X.-G. He, G.~C. Joshi, H.~Lew, and R.~R. Volkas, {\it {Simplest Z-prime
  model}},  {\em Phys. Rev.} {\bf D44} (1991) 2118--2132.

\bibitem{Heeck:2011wj}
J.~Heeck and W.~Rodejohann, {\it {Gauged $L_\mu - L_\tau$ Symmetry at the
  Electroweak Scale}},  {\em Phys. Rev.} {\bf D84} (2011) 075007,
  [\href{http://xxx.lanl.gov/abs/1107.5238}{{\tt 1107.5238}}].

\bibitem{Aaij:2013qta}
{\bf LHCb} Collaboration, R.~Aaij {\em et.~al.}, {\it {Measurement of
  Form-Factor-Independent Observables in the Decay $B^{0} \to K^{*0} \mu^+
  \mu^-$}},  {\em Phys. Rev. Lett.} {\bf 111} (2013) 191801,
  [\href{http://xxx.lanl.gov/abs/1308.1707}{{\tt 1308.1707}}].

\bibitem{Glashow:2014iga}
S.~L. Glashow, D.~Guadagnoli, and K.~Lane, {\it {Lepton Flavor Violation in $B$
  Decays?}},  {\em Phys. Rev. Lett.} {\bf 114} (2015) 091801,
  [\href{http://xxx.lanl.gov/abs/1411.0565}{{\tt 1411.0565}}].

\bibitem{Crivellin:2015mga}
A.~Crivellin, G.~D'Ambrosio, and J.~Heeck, {\it {Explaining
  $h\to\mu^\pm\tau^\mp$, $B\to K^* \mu^+\mu^-$ and $B\to K \mu^+\mu^-/B\to K
  e^+e^-$ in a two-Higgs-doublet model with gauged $L_\mu-L_\tau$}},  {\em
  Phys. Rev. Lett.} {\bf 114} (2015) 151801,
  [\href{http://xxx.lanl.gov/abs/1501.00993}{{\tt 1501.00993}}].

\bibitem{Varzielas:2015iva}
I.~de~Medeiros~Varzielas and G.~Hiller, {\it {Clues for flavor from rare lepton
  and quark decays}},  {\em JHEP} {\bf 06} (2015) 072,
  [\href{http://xxx.lanl.gov/abs/1503.01084}{{\tt 1503.01084}}].

\bibitem{Crivellin:2015lwa}
A.~Crivellin, G.~D'Ambrosio, and J.~Heeck, {\it {Addressing the LHC flavor
  anomalies with horizontal gauge symmetries}},  {\em Phys. Rev.} {\bf D91}
  (2015), no.~7 075006, [\href{http://xxx.lanl.gov/abs/1503.03477}{{\tt
  1503.03477}}].

\bibitem{Niehoff:2015bfa}
C.~Niehoff, P.~Stangl, and D.~M. Straub, {\it {Violation of lepton flavour
  universality in composite Higgs models}},  {\em Phys. Lett.} {\bf B747}
  (2015) 182--186, [\href{http://xxx.lanl.gov/abs/1503.03865}{{\tt
  1503.03865}}].

\bibitem{Sierra:2015fma}
D.~Aristizabal~Sierra, F.~Staub, and A.~Vicente, {\it {Shedding light on the
  $b\to s$ anomalies with a dark sector}},  {\em Phys. Rev.} {\bf D92} (2015),
  no.~1 015001, [\href{http://xxx.lanl.gov/abs/1503.06077}{{\tt 1503.06077}}].

\bibitem{Celis:2015ara}
A.~Celis, J.~Fuentes-Martin, M.~Jung, and H.~Serodio, {\it {Family nonuniversal
  $Z^\prime$ models with protected flavor-changing interactions}},  {\em Phys.
  Rev.} {\bf D92} (2015), no.~1 015007,
  [\href{http://xxx.lanl.gov/abs/1505.03079}{{\tt 1505.03079}}].

\bibitem{Greljo:2015mma}
A.~Greljo, G.~Isidori, and D.~Marzocca, {\it {On the breaking of Lepton Flavor
  Universality in B decays}},  {\em JHEP} {\bf 07} (2015) 142,
  [\href{http://xxx.lanl.gov/abs/1506.01705}{{\tt 1506.01705}}].

\bibitem{Belanger:2015nma}
G.~Bélanger, C.~Delaunay, and S.~Westhoff, {\it {A Dark Matter Relic From Muon
  Anomalies}},  {\em Phys. Rev.} {\bf D92} (2015) 055021,
  [\href{http://xxx.lanl.gov/abs/1507.06660}{{\tt 1507.06660}}].

\bibitem{Falkowski:2015zwa}
A.~Falkowski, M.~Nardecchia, and R.~Ziegler, {\it {Lepton Flavor
  Non-Universality in B-meson Decays from a U(2) Flavor Model}},  {\em JHEP}
  {\bf 11} (2015) 173, [\href{http://xxx.lanl.gov/abs/1509.01249}{{\tt
  1509.01249}}].

\bibitem{Allanach:2015gkd}
B.~Allanach, F.~S. Queiroz, A.~Strumia, and S.~Sun, {\it {Z' models for the
  LHCb and g-2 muon anomalies}},
  \href{http://xxx.lanl.gov/abs/1511.07447}{{\tt 1511.07447}}.

\bibitem{Fuyuto:2015gmk}
K.~Fuyuto, W.-S. Hou, and M.~Kohda, {\it {Z'-induced FCNC decays of top,
  beauty, and strange quarks}},  {\em Phys. Rev.} {\bf D93} (2016), no.~5
  054021, [\href{http://xxx.lanl.gov/abs/1512.09026}{{\tt 1512.09026}}].

\bibitem{Becirevic:2016zri}
D.~Becirevic, O.~Sumensari, and R.~Zukanovich~Funchal, {\it {Lepton flavor
  violation in exclusive $b\rightarrow s$ decays}},  {\em Eur. Phys. J.} {\bf
  C76} (2016), no.~3 134, [\href{http://xxx.lanl.gov/abs/1602.00881}{{\tt
  1602.00881}}].

\bibitem{Altmannshofer:2016oaq}
W.~Altmannshofer, M.~Carena, and A.~Crivellin, {\it {A $L_\mu - L_\tau$ Theory
  of Higgs Flavor Violation and $(g-2)_\mu$}},
  \href{http://xxx.lanl.gov/abs/1604.08221}{{\tt 1604.08221}}.

\bibitem{Boucenna:2016qad}
S.~M. Boucenna, A.~Celis, J.~Fuentes-Martin, A.~Vicente, and J.~Virto, {\it
  {Phenomenology of an $SU(2) \times SU(2) \times U(1)$ model with
  lepton-flavour non-universality}},
  \href{http://xxx.lanl.gov/abs/1608.01349}{{\tt 1608.01349}}.

\bibitem{Megias:2016bde}
E.~Megias, G.~Panico, O.~Pujolas, and M.~Quiros, {\it {A Natural origin for the
  LHCb anomalies}},  \href{http://xxx.lanl.gov/abs/1608.02362}{{\tt
  1608.02362}}.

\bibitem{Aaij:2015oid}
{\bf LHCb} Collaboration, R.~Aaij {\em et.~al.}, {\it {Angular analysis of the
  $B^{0} \to K^{*0} \mu^{+} \mu^{-}$ decay using 3 fb$^{-1}$ of integrated
  luminosity}},  {\em JHEP} {\bf 02} (2016) 104,
  [\href{http://xxx.lanl.gov/abs/1512.04442}{{\tt 1512.04442}}].

\bibitem{Abdesselam:2016llu}
{\bf Belle} Collaboration, A.~Abdesselam {\em et.~al.}, {\it {Angular analysis
  of $B^0 \to K^\ast(892)^0 \ell^+ \ell^-$}},  in {\em {LHC Ski 2016: A First
  Discussion of 13 TeV Results Obergurgl, Austria, April 10-15, 2016}}, 2016.
\newblock \href{http://xxx.lanl.gov/abs/1604.04042}{{\tt 1604.04042}}.

\bibitem{Aaij:2015esa}
{\bf LHCb} Collaboration, R.~Aaij {\em et.~al.}, {\it {Angular analysis and
  differential branching fraction of the decay $B^0_s\to\phi\mu^+\mu^-$}},
  {\em JHEP} {\bf 09} (2015) 179,
  [\href{http://xxx.lanl.gov/abs/1506.08777}{{\tt 1506.08777}}].

\bibitem{Aaij:2014ora}
{\bf LHCb} Collaboration, R.~Aaij {\em et.~al.}, {\it {Test of lepton
  universality using $B^{+}\rightarrow K^{+}\ell^{+}\ell^{-}$ decays}},  {\em
  Phys. Rev. Lett.} {\bf 113} (2014) 151601,
  [\href{http://xxx.lanl.gov/abs/1406.6482}{{\tt 1406.6482}}].

\bibitem{Altmannshofer:2014rta}
W.~Altmannshofer and D.~M. Straub, {\it {New physics in $b\rightarrow s$
  transitions after LHC run 1}},  {\em Eur. Phys. J.} {\bf C75} (2015), no.~8
  382, [\href{http://xxx.lanl.gov/abs/1411.3161}{{\tt 1411.3161}}].

\bibitem{Altmannshofer:2015sma}
W.~Altmannshofer and D.~M. Straub, {\it {Implications of $b\to s$
  measurements}},  in {\em {Proceedings, 50th Rencontres de Moriond Electroweak
  interactions and unified theories}}, pp.~333--338, 2015.
\newblock \href{http://xxx.lanl.gov/abs/1503.06199}{{\tt 1503.06199}}.

\bibitem{Descotes-Genon:2015uva}
S.~Descotes-Genon, L.~Hofer, J.~Matias, and J.~Virto, {\it {Global analysis of
  $b\to s\ell\ell$ anomalies}},  \href{http://xxx.lanl.gov/abs/1510.04239}{{\tt
  1510.04239}}.

\bibitem{Hurth:2016fbr}
T.~Hurth, F.~Mahmoudi, and S.~Neshatpour, {\it {On the anomalies in the latest
  LHCb data}},  {\em Nucl. Phys.} {\bf B909} (2016) 737--777,
  [\href{http://xxx.lanl.gov/abs/1603.00865}{{\tt 1603.00865}}].

\bibitem{Ruegg:2003ps}
H.~Ruegg and M.~Ruiz-Altaba, {\it {The Stueckelberg field}},  {\em Int. J. Mod.
  Phys.} {\bf A19} (2004) 3265--3348,
  [\href{http://xxx.lanl.gov/abs/hep-th/0304245}{{\tt hep-th/0304245}}].

\bibitem{Feldman:2006wb}
D.~Feldman, Z.~Liu, and P.~Nath, {\it {The Stueckelberg $Z^\prime$ at the LHC:
  Discovery Potential, Signature Spaces and Model Discrimination}},  {\em JHEP}
  {\bf 11} (2006) 007, [\href{http://xxx.lanl.gov/abs/hep-ph/0606294}{{\tt
  hep-ph/0606294}}].

\bibitem{Baek:2008nz}
S.~Baek and P.~Ko, {\it {Phenomenology of $U(1)_{L_\mu-L_\tau}$ charged dark
  matter at PAMELA and colliders}},  {\em JCAP} {\bf 0910} (2009) 011,
  [\href{http://xxx.lanl.gov/abs/0811.1646}{{\tt 0811.1646}}].

\bibitem{Bi:2009uj}
X.-J. Bi, X.-G. He, and Q.~Yuan, {\it {Parameters in a class of leptophilic
  models from PAMELA, ATIC and FERMI}},  {\em Phys. Lett.} {\bf B678} (2009)
  168--173, [\href{http://xxx.lanl.gov/abs/0903.0122}{{\tt 0903.0122}}].

\bibitem{Queiroz:2014zfa}
F.~S. Queiroz and W.~Shepherd, {\it {New Physics Contributions to the Muon
  Anomalous Magnetic Moment: A Numerical Code}},  {\em Phys. Rev.} {\bf D89}
  (2014), no.~9 095024, [\href{http://xxx.lanl.gov/abs/1403.2309}{{\tt
  1403.2309}}].

\bibitem{Kim:2015fpa}
J.-C. Park, S.~C. Park, and J.~Kim, {\it {Galactic center GeV gamma-ray excess
  from dark matter with gauged lepton numbers}},  {\em Phys. Lett.} {\bf B752}
  (2016) 59--65, [\href{http://xxx.lanl.gov/abs/1505.04620}{{\tt 1505.04620}}].

\bibitem{Baek:2015fea}
S.~Baek, {\it {Dark matter and muon $(g-2)$ in local
  $U(1)_{L_\mu-L_\tau}$-extended Ma Model}},  {\em Phys. Lett.} {\bf B756}
  (2016) 1--5, [\href{http://xxx.lanl.gov/abs/1510.02168}{{\tt 1510.02168}}].

\bibitem{Kile:2014jea}
J.~Kile, A.~Kobach, and A.~Soni, {\it {Lepton-Flavored Dark Matter}},  {\em
  Phys. Lett.} {\bf B744} (2015) 330--338,
  [\href{http://xxx.lanl.gov/abs/1411.1407}{{\tt 1411.1407}}].

\bibitem{Patra:2016shz}
S.~Patra, S.~Rao, N.~Sahoo, and N.~Sahu, {\it {Gauged $U(1)_{L_\mu - L_\tau}$
  model in light of muon $g-2$ anomaly, neutrino mass and dark matter
  phenomenology}},  \href{http://xxx.lanl.gov/abs/1607.04046}{{\tt
  1607.04046}}.

\bibitem{Biswas:2016yan}
A.~Biswas, S.~Choubey, and S.~Khan, {\it {Neutrino Mass, Dark Matter and
  Anomalous Magnetic Moment of Muon in a $U(1)_{L_{\mu}-L_{\tau}}$ Model}},
  \href{http://xxx.lanl.gov/abs/1608.04194}{{\tt 1608.04194}}.

\bibitem{Celis:2016ayl}
A.~Celis, W.-Z. Feng, and M.~Vollmann, {\it {Dirac Dark Matter and $b \to s
  \ell^+ \ell^-$ with $U(1)$ gauge symmetry}},
  \href{http://xxx.lanl.gov/abs/1608.03894}{{\tt 1608.03894}}.

\bibitem{Altmannshofer:2014pba}
W.~Altmannshofer, S.~Gori, M.~Pospelov, and I.~Yavin, {\it {Neutrino Trident
  Production: A Powerful Probe of New Physics with Neutrino Beams}},  {\em
  Phys. Rev. Lett.} {\bf 113} (2014) 091801,
  [\href{http://xxx.lanl.gov/abs/1406.2332}{{\tt 1406.2332}}].

\bibitem{Mishra:1991bv}
{\bf CCFR} Collaboration, S.~R. Mishra {\em et.~al.}, {\it {Neutrino tridents
  and W Z interference}},  {\em Phys. Rev. Lett.} {\bf 66} (1991) 3117--3120.

\bibitem{TheATLAScollaboration:2013nha}
{\bf ATLAS} Collaboration, {\it {ATLAS measurements of the 7 and 8 TeV cross
  sections for $Z\rightarrow 4\ell$ in pp collisions}}, .

\bibitem{Aad:2014wra}
{\bf ATLAS} Collaboration, G.~Aad {\em et.~al.}, {\it {Measurements of
  Four-Lepton Production at the Z Resonance in pp Collisions at $\sqrt s=$7 and
  8 TeV with ATLAS}},  {\em Phys. Rev. Lett.} {\bf 112} (2014), no.~23 231806,
  [\href{http://xxx.lanl.gov/abs/1403.5657}{{\tt 1403.5657}}].

\bibitem{CMS:2016vvl}
{\bf CMS} Collaboration, {\it {Measurement of the ZZ production cross section
  and $\mathrm{Z} \to \ell\ell\ell'\ell'$ branching fraction in pp collisions
  at $\sqrt{s}=13~\mathrm{TeV}$}}, .

\bibitem{Elahi:2015vzh}
F.~Elahi and A.~Martin, {\it {Constraints on $L_\mu - L_\tau$ interactions at
  the LHC and beyond}},  {\em Phys. Rev.} {\bf D93} (2016), no.~1 015022,
  [\href{http://xxx.lanl.gov/abs/1511.04107}{{\tt 1511.04107}}].

\bibitem{TheBABAR:2016rlg}
{\bf BaBar} Collaboration, J.~P. Lees {\em et.~al.}, {\it {Search for a muonic
  dark force at BABAR}},  \href{http://xxx.lanl.gov/abs/1606.03501}{{\tt
  1606.03501}}.

\bibitem{Blake:2016olu}
T.~Blake, G.~Lanfranchi, and D.~M. Straub, {\it {Rare $B$ Decays as Tests of
  the Standard Model}},  \href{http://xxx.lanl.gov/abs/1606.00916}{{\tt
  1606.00916}}.

\bibitem{Aaij:2015tna}
{\bf LHCb} Collaboration, R.~Aaij {\em et.~al.}, {\it {Search for hidden-sector
  bosons in $B^0 \!\to K^{*0}\mu^+\mu^-$ decays}},  {\em Phys. Rev. Lett.} {\bf
  115} (2015), no.~16 161802, [\href{http://xxx.lanl.gov/abs/1508.04094}{{\tt
  1508.04094}}].

\bibitem{Buchalla:1995vs}
G.~Buchalla, A.~J. Buras, and M.~E. Lautenbacher, {\it {Weak decays beyond
  leading logarithms}},  {\em Rev. Mod. Phys.} {\bf 68} (1996) 1125--1144,
  [\href{http://xxx.lanl.gov/abs/hep-ph/9512380}{{\tt hep-ph/9512380}}].

\bibitem{Charles:2015gya}
J.~Charles {\em et.~al.}, {\it {Current status of the Standard Model CKM fit
  and constraints on $\Delta F=2$ New Physics}},  {\em Phys. Rev.} {\bf D91}
  (2015), no.~7 073007, [\href{http://xxx.lanl.gov/abs/1501.05013}{{\tt
  1501.05013}}].

\bibitem{Bazavov:2016nty}
{\bf Fermilab Lattice, MILC} Collaboration, A.~Bazavov {\em et.~al.}, {\it
  {$B^0_{(s)}$-Mixing Matrix Elements from Lattice QCD for the Standard Model
  and Beyond}},  \href{http://xxx.lanl.gov/abs/1602.03560}{{\tt 1602.03560}}.

\bibitem{Mizukoshi:2010ky}
J.~K. Mizukoshi, C.~A. de~S.~Pires, F.~S. Queiroz, and P.~S. Rodrigues~da
  Silva, {\it {WIMPs in a 3-3-1 model with heavy Sterile neutrinos}},  {\em
  Phys. Rev.} {\bf D83} (2011) 065024,
  [\href{http://xxx.lanl.gov/abs/1010.4097}{{\tt 1010.4097}}].

\bibitem{Profumo:2013sca}
S.~Profumo and F.~S. Queiroz, {\it {Constraining the $Z'$ mass in 331 models
  using direct dark matter detection}},  {\em Eur. Phys. J.} {\bf C74} (2014),
  no.~7 2960, [\href{http://xxx.lanl.gov/abs/1307.7802}{{\tt 1307.7802}}].

\bibitem{Alves:2013tqa}
A.~Alves, S.~Profumo, and F.~S. Queiroz, {\it {The dark $Z^{'}$ portal: direct,
  indirect and collider searches}},  {\em JHEP} {\bf 04} (2014) 063,
  [\href{http://xxx.lanl.gov/abs/1312.5281}{{\tt 1312.5281}}].

\bibitem{Arcadi:2014lta}
G.~Arcadi, Y.~Mambrini, and F.~Richard, {\it {Z-portal dark matter}},  {\em
  JCAP} {\bf 1503} (2015) 018, [\href{http://xxx.lanl.gov/abs/1411.2985}{{\tt
  1411.2985}}].

\bibitem{Buchmueller:2014yoa}
O.~Buchmueller, M.~J. Dolan, S.~A. Malik, and C.~McCabe, {\it {Characterising
  dark matter searches at colliders and direct detection experiments: Vector
  mediators}},  {\em JHEP} {\bf 01} (2015) 037,
  [\href{http://xxx.lanl.gov/abs/1407.8257}{{\tt 1407.8257}}].

\bibitem{Cline:2014dwa}
J.~M. Cline, G.~Dupuis, Z.~Liu, and W.~Xue, {\it {The windows for kinetically
  mixed Z'-mediated dark matter and the galactic center gamma ray excess}},
  {\em JHEP} {\bf 08} (2014) 131,
  [\href{http://xxx.lanl.gov/abs/1405.7691}{{\tt 1405.7691}}].

\bibitem{Fairbairn:2014aqa}
M.~Fairbairn and J.~Heal, {\it {Complementarity of dark matter searches at
  resonance}},  {\em Phys. Rev.} {\bf D90} (2014), no.~11 115019,
  [\href{http://xxx.lanl.gov/abs/1406.3288}{{\tt 1406.3288}}].

\bibitem{Lebedev:2014bba}
O.~Lebedev and Y.~Mambrini, {\it {Axial dark matter: The case for an invisible
  $Z′$}},  {\em Phys. Lett.} {\bf B734} (2014) 350--353,
  [\href{http://xxx.lanl.gov/abs/1403.4837}{{\tt 1403.4837}}].

\bibitem{deSimone:2014pda}
A.~De~Simone, G.~F. Giudice, and A.~Strumia, {\it {Benchmarks for Dark Matter
  Searches at the LHC}},  {\em JHEP} {\bf 06} (2014) 081,
  [\href{http://xxx.lanl.gov/abs/1402.6287}{{\tt 1402.6287}}].

\bibitem{Okada:2016gsh}
N.~Okada and S.~Okada, {\it {$Z^\prime_{BL}$ portal dark matter and LHC Run-2
  results}},  \href{http://xxx.lanl.gov/abs/1601.07526}{{\tt 1601.07526}}.

\bibitem{Alves:2015mua}
A.~Alves, A.~Berlin, S.~Profumo, and F.~S. Queiroz, {\it {Dirac-fermionic dark
  matter in U(1)$_{X}$ models}},  {\em JHEP} {\bf 10} (2015) 076,
  [\href{http://xxx.lanl.gov/abs/1506.06767}{{\tt 1506.06767}}].

\bibitem{Kahlhoefer:2015bea}
F.~Kahlhoefer, K.~Schmidt-Hoberg, T.~Schwetz, and S.~Vogl, {\it {Implications
  of unitarity and gauge invariance for simplified dark matter models}},  {\em
  JHEP} {\bf 02} (2016) 016, [\href{http://xxx.lanl.gov/abs/1510.02110}{{\tt
  1510.02110}}]. [JHEP02,016(2016)].

\bibitem{Duerr:2015wfa}
M.~Duerr, P.~Fileviez~Perez, and J.~Smirnov, {\it {Simplified Dirac Dark Matter
  Models}},  \href{http://xxx.lanl.gov/abs/1506.05107}{{\tt 1506.05107}}.

\bibitem{Brennan:2016xjh}
A.~J. Brennan, M.~F. McDonald, J.~Gramling, and T.~D. Jacques, {\it {Collide
  and Conquer: Constraints on Simplified Dark Matter Models using Mono-X
  Collider Searches}},  \href{http://xxx.lanl.gov/abs/1603.01366}{{\tt
  1603.01366}}.

\bibitem{Jacques:2016dqz}
T.~Jacques, A.~Katz, E.~Morgante, D.~Racco, M.~Rameez, and A.~Riotto, {\it
  {Complementarity of DM Searches in a Consistent Simplified Model: the Case of
  Z'}},  \href{http://xxx.lanl.gov/abs/1605.06513}{{\tt 1605.06513}}.

\bibitem{Englert:2016joy}
C.~Englert, M.~McCullough, and M.~Spannowsky, {\it {S-Channel Dark Matter
  Simplified Models and Unitarity}},
  \href{http://xxx.lanl.gov/abs/1604.07975}{{\tt 1604.07975}}.

\bibitem{D'Eramo:2016atc}
F.~D'Eramo, B.~J. Kavanagh, and P.~Panci, {\it {You can hide but you have to
  run: direct detection with vector mediators}},
  \href{http://xxx.lanl.gov/abs/1605.04917}{{\tt 1605.04917}}.

\bibitem{Klasen:2016qux}
M.~Klasen, F.~Lyonnet, and F.~S. Queiroz, {\it {NLO+NLL Collider Bounds, Dirac
  Fermion and Scalar Dark Matter in the B-L Model}},
  \href{http://xxx.lanl.gov/abs/1607.06468}{{\tt 1607.06468}}.

\bibitem{Belanger:2006is}
G.~Belanger, F.~Boudjema, A.~Pukhov, and A.~Semenov, {\it {MicrOMEGAs 2.0: A
  Program to calculate the relic density of dark matter in a generic model}},
  {\em Comput. Phys. Commun.} {\bf 176} (2007) 367--382,
  [\href{http://xxx.lanl.gov/abs/hep-ph/0607059}{{\tt hep-ph/0607059}}].

\bibitem{Belanger:2008sj}
G.~Belanger, F.~Boudjema, A.~Pukhov, and A.~Semenov, {\it {Dark matter direct
  detection rate in a generic model with micrOMEGAs 2.2}},  {\em Comput. Phys.
  Commun.} {\bf 180} (2009) 747--767,
  [\href{http://xxx.lanl.gov/abs/0803.2360}{{\tt 0803.2360}}].

\bibitem{Beltran:2008xg}
M.~Beltran, D.~Hooper, E.~W. Kolb, and Z.~C. Krusberg, {\it {Deducing the
  nature of dark matter from direct and indirect detection experiments in the
  absence of collider signatures of new physics}},  {\em Phys. Rev.} {\bf D80}
  (2009) 043509, [\href{http://xxx.lanl.gov/abs/0808.3384}{{\tt 0808.3384}}].

\bibitem{Beltran:2010ww}
M.~Beltran, D.~Hooper, E.~W. Kolb, Z.~A.~C. Krusberg, and T.~M.~P. Tait, {\it
  {Maverick dark matter at colliders}},  {\em JHEP} {\bf 09} (2010) 037,
  [\href{http://xxx.lanl.gov/abs/1002.4137}{{\tt 1002.4137}}].

\bibitem{Ade:2015xua}
{\bf Planck} Collaboration, P.~A.~R. Ade {\em et.~al.}, {\it {Planck 2015
  results. XIII. Cosmological parameters}},
  \href{http://xxx.lanl.gov/abs/1502.01589}{{\tt 1502.01589}}.

\bibitem{Bergstrom:2013jra}
L.~Bergstrom, T.~Bringmann, I.~Cholis, D.~Hooper, and C.~Weniger, {\it {New
  limits on dark matter annihilation from AMS cosmic ray positron data}},  {\em
  Phys. Rev. Lett.} {\bf 111} (2013) 171101,
  [\href{http://xxx.lanl.gov/abs/1306.3983}{{\tt 1306.3983}}].

\bibitem{Ibarra:2013zia}
A.~Ibarra, A.~S. Lamperstorfer, and J.~Silk, {\it {Dark matter annihilations
  and decays after the AMS-02 positron measurements}},  {\em Phys. Rev.} {\bf
  D89} (2014), no.~6 063539, [\href{http://xxx.lanl.gov/abs/1309.2570}{{\tt
  1309.2570}}].

\bibitem{Kopp:2013eka}
J.~Kopp, {\it {Constraints on dark matter annihilation from AMS-02 results}},
  {\em Phys. Rev.} {\bf D88} (2013) 076013,
  [\href{http://xxx.lanl.gov/abs/1304.1184}{{\tt 1304.1184}}].

\bibitem{Lu:2015pta}
B.-Q. Lu and H.-S. Zong, {\it {Limits on dark matter from AMS-02 antiproton and
  positron fraction data}},  {\em Phys. Rev.} {\bf D93} (2016), no.~10 103517,
  [\href{http://xxx.lanl.gov/abs/1510.04032}{{\tt 1510.04032}}].

\bibitem{DiMauro:2015jxa}
M.~Di~Mauro, F.~Donato, N.~Fornengo, and A.~Vittino, {\it {Dark matter vs.
  astrophysics in the interpretation of AMS-02 electron and positron data}},
  {\em JCAP} {\bf 1605} (2016), no.~05 031,
  [\href{http://xxx.lanl.gov/abs/1507.07001}{{\tt 1507.07001}}].

\bibitem{Slatyer:2009yq}
T.~R. Slatyer, N.~Padmanabhan, and D.~P. Finkbeiner, {\it {CMB Constraints on
  WIMP Annihilation: Energy Absorption During the Recombination Epoch}},  {\em
  Phys. Rev.} {\bf D80} (2009) 043526,
  [\href{http://xxx.lanl.gov/abs/0906.1197}{{\tt 0906.1197}}].

\bibitem{Finkbeiner:2011dx}
D.~P. Finkbeiner, S.~Galli, T.~Lin, and T.~R. Slatyer, {\it {Searching for Dark
  Matter in the CMB: A Compact Parameterization of Energy Injection from New
  Physics}},  {\em Phys. Rev.} {\bf D85} (2012) 043522,
  [\href{http://xxx.lanl.gov/abs/1109.6322}{{\tt 1109.6322}}].

\bibitem{Galli:2011rz}
S.~Galli, F.~Iocco, G.~Bertone, and A.~Melchiorri, {\it {Updated CMB
  constraints on Dark Matter annihilation cross-sections}},  {\em Phys. Rev.}
  {\bf D84} (2011) 027302, [\href{http://xxx.lanl.gov/abs/1106.1528}{{\tt
  1106.1528}}].

\bibitem{Weniger:2013hja}
C.~Weniger, P.~D. Serpico, F.~Iocco, and G.~Bertone, {\it {CMB bounds on dark
  matter annihilation: Nucleon energy-losses after recombination}},  {\em Phys.
  Rev.} {\bf D87} (2013), no.~12 123008,
  [\href{http://xxx.lanl.gov/abs/1303.0942}{{\tt 1303.0942}}].

\bibitem{Lopez-Honorez:2013lcm}
L.~Lopez-Honorez, O.~Mena, S.~Palomares-Ruiz, and A.~C. Vincent, {\it
  {Constraints on dark matter annihilation from CMB observations before
  Planck}},  {\em JCAP} {\bf 1307} (2013) 046,
  [\href{http://xxx.lanl.gov/abs/1303.5094}{{\tt 1303.5094}}].

\bibitem{Madhavacheril:2013cna}
M.~S. Madhavacheril, N.~Sehgal, and T.~R. Slatyer, {\it {Current Dark Matter
  Annihilation Constraints from CMB and Low-Redshift Data}},  {\em Phys. Rev.}
  {\bf D89} (2014) 103508, [\href{http://xxx.lanl.gov/abs/1310.3815}{{\tt
  1310.3815}}].

\bibitem{Galli:2013dna}
S.~Galli, T.~R. Slatyer, M.~Valdes, and F.~Iocco, {\it {Systematic
  Uncertainties In Constraining Dark Matter Annihilation From The Cosmic
  Microwave Background}},  {\em Phys. Rev.} {\bf D88} (2013) 063502,
  [\href{http://xxx.lanl.gov/abs/1306.0563}{{\tt 1306.0563}}].

\bibitem{Slatyer:2015kla}
T.~R. Slatyer, {\it {Indirect Dark Matter Signatures in the Cosmic Dark Ages
  II. Ionization, Heating and Photon Production from Arbitrary Energy
  Injections}},  {\em Phys. Rev.} {\bf D93} (2016), no.~2 023521,
  [\href{http://xxx.lanl.gov/abs/1506.03812}{{\tt 1506.03812}}].

\bibitem{Slatyer:2015jla}
T.~R. Slatyer, {\it {Indirect dark matter signatures in the cosmic dark ages.
  I. Generalizing the bound on s-wave dark matter annihilation from Planck
  results}},  {\em Phys. Rev.} {\bf D93} (2016), no.~2 023527,
  [\href{http://xxx.lanl.gov/abs/1506.03811}{{\tt 1506.03811}}].

\bibitem{Adrian-Martinez:2016gti}
{\bf ANTARES} Collaboration, S.~Adrian-Martinez {\em et.~al.}, {\it {Limits on
  Dark Matter Annihilation in the Sun using the ANTARES Neutrino Telescope}},
  {\em Phys. Lett.} {\bf B759} (2016) 69--74,
  [\href{http://xxx.lanl.gov/abs/1603.02228}{{\tt 1603.02228}}].

\bibitem{Adrian-Martinez:2015wey}
{\bf ANTARES} Collaboration, S.~Adrian-Martinez {\em et.~al.}, {\it {Search of
  Dark Matter Annihilation in the Galactic Centre using the ANTARES Neutrino
  Telescope}},  {\em JCAP} {\bf 1510} (2015), no.~10 068,
  [\href{http://xxx.lanl.gov/abs/1505.04866}{{\tt 1505.04866}}].

\bibitem{Queiroz:2016zwd}
F.~S. Queiroz, C.~E. Yaguna, and C.~Weniger, {\it {Gamma-ray Limits on Neutrino
  Lines}},  {\em JCAP} {\bf 1605} (2016), no.~05 050,
  [\href{http://xxx.lanl.gov/abs/1602.05966}{{\tt 1602.05966}}].

\bibitem{Griest:1990kh}
K.~Griest and D.~Seckel, {\it {Three exceptions in the calculation of relic
  abundances}},  {\em Phys.Rev.} {\bf D43} (1991) 3191--3203.

\bibitem{Edsjo:1997bg}
J.~Edsjo and P.~Gondolo, {\it {Neutralino relic density including
  coannihilations}},  {\em Phys. Rev.} {\bf D56} (1997) 1879--1894,
  [\href{http://xxx.lanl.gov/abs/hep-ph/9704361}{{\tt hep-ph/9704361}}].

\bibitem{Kopp:2009et}
J.~Kopp, V.~Niro, T.~Schwetz, and J.~Zupan, {\it {DAMA/LIBRA and leptonically
  interacting Dark Matter}},  {\em Phys. Rev.} {\bf D80} (2009) 083502,
  [\href{http://xxx.lanl.gov/abs/0907.3159}{{\tt 0907.3159}}].

\bibitem{Akerib:2016vxi}
D.~S. Akerib {\em et.~al.}, {\it {Results from a search for dark matter in LUX
  with 332 live days of exposure}},
  \href{http://xxx.lanl.gov/abs/1608.07648}{{\tt 1608.07648}}.

\bibitem{Akerib:2013tjd}
{\bf LUX} Collaboration, D.~S. Akerib {\em et.~al.}, {\it {First results from
  the LUX dark matter experiment at the Sanford Underground Research
  Facility}},  {\em Phys. Rev. Lett.} {\bf 112} (2014) 091303,
  [\href{http://xxx.lanl.gov/abs/1310.8214}{{\tt 1310.8214}}].

\bibitem{Akerib:2015rjg}
{\bf LUX} Collaboration, D.~S. Akerib {\em et.~al.}, {\it {Improved Limits on
  Scattering of Weakly Interacting Massive Particles from Reanalysis of 2013
  LUX Data}},  {\em Phys. Rev. Lett.} {\bf 116} (2016), no.~16 161301,
  [\href{http://xxx.lanl.gov/abs/1512.03506}{{\tt 1512.03506}}].

\bibitem{Tan:2016zwf}
{\bf PandaX-II} Collaboration, A.~Tan {\em et.~al.}, {\it {Dark Matter Results
  from First 98.7-day Data of PandaX-II Experiment}},
  \href{http://xxx.lanl.gov/abs/1607.07400}{{\tt 1607.07400}}.

\bibitem{Aprile:2015uzo}
{\bf XENON} Collaboration, E.~Aprile {\em et.~al.}, {\it {Physics reach of the
  XENON1T dark matter experiment}},  {\em JCAP} {\bf 1604} (2016), no.~04 027,
  [\href{http://xxx.lanl.gov/abs/1512.07501}{{\tt 1512.07501}}].

\bibitem{Akerib:2015cja}
{\bf LZ} Collaboration, D.~S. Akerib {\em et.~al.}, {\it {LUX-ZEPLIN (LZ)
  Conceptual Design Report}},  \href{http://xxx.lanl.gov/abs/1509.02910}{{\tt
  1509.02910}}.

\end{thebibliography}\endgroup

\end{document}